\documentclass[letterpaper]{emulateapj}
\usepackage{amsmath}
\usepackage{amssymb}
\usepackage{latexsym}
\usepackage{textcomp}
\usepackage{graphicx}
\usepackage{subfigure}
\usepackage{lscape}
\usepackage{epsfig}
\usepackage{natbib}

\newcommand{\ha}{\textrm{H}\ensuremath{\alpha}}
\newcommand{\hb}{\textrm{H}\ensuremath{\beta}}

\newcommand{\nii}{[\textrm{N}\,\textsc{ii}]}
\newcommand{\oiii}{[\textrm{O}\,\textsc{iii}]}
\newcommand{\hii}{\textrm{H}\,\textsc{ii}}
\newcommand{\sii}{[\textrm{S}\,\textsc{ii}]}

\newcommand{\oiiilam}{[\textrm{O}\,\textsc{iii}]\,\ensuremath{\lambda5007}}
\newcommand{\niilam}{[\textrm{N}\,\textsc{ii}]\,\ensuremath{\lambda6584}} 

\newcommand{\oiilam}{[\textrm{O}\,\textsc{ii}]\,\ensuremath{\lambda3727}}
\newcommand{\oilam}{[\textrm{O}\,\textsc{i}]\,\ensuremath{\lambda6300}}
\newcommand{\siilam}{[\textrm{S}\,\textsc{ii}]\,\ensuremath{\lambda\lambda6717,6731}}

\newcommand{\neiiilam}{[\textrm{Ne}\,\textsc{iii}]\,\ensuremath{\lambda3869}} 

\newcommand{\nevlam}{[\textrm{Ne}\,\textsc{v}]\,\ensuremath{\lambda3425}} 
\newcommand{\nevmir}{[\textrm{Ne}\,\textsc{v}]\,\ensuremath{\lambda14\mu}{\rm m}} 
\newcommand{\oivlam}{[\textrm{O}\,\textsc{iv}]\,\ensuremath{\lambda25.9\mu}{\rm m}}

\slugcomment{To appear in the ApJ}

\shorttitle{AGN Diagnostic at $z > 0.3$}
\shortauthors{Juneau et al.}

\begin{document}

\title{A New Diagnostic of Active Galactic Nuclei: Revealing Highly-Absorbed Systems at Redshift$>0.3$}

\author{\sc St\'{e}phanie Juneau} 
\affil{Steward Observatory, University of Arizona, Tucson, AZ 85721; sjuneau@as.arizona.edu}

\author{\sc Mark Dickinson} 
\affil{National Optical Astronomy Observatory, 950 North Cherry Avenue, Tucson, AZ 85719; med@noao.edu}

\author{\sc David M. Alexander} 
\affil{Department of Physics, Durham University, Durham DH1 3LE, UK; d.m.alexander@durham.ac.uk}

\and

\author{\sc Samir Salim} 
\affil{Department of Astronomy, Indiana University, Bloomington, IN 47404; salims@indiana.edu}

\begin{abstract}

We introduce the Mass-Excitation (MEx) diagnostic to identify active 
galactic nuclei (AGNs) in galaxies at intermediate redshift. 
In the absence of near-infrared spectroscopy, necessary to use 
traditional nebular line diagrams at $z>0.4$, we demonstrate that 
combining \oiiilam/\hb\ and stellar mass successfully distinguishes 
between star formation and AGN emission.  
The MEx classification scheme relies on a novel probabilistic 
approach splitting galaxies into sub-categories with more 
confidence than alternative high-$z$ diagnostic diagrams.  It recognizes 
that galaxies near empirical boundaries on traditional diagrams have an 
uncertain classification and thus a non-zero probability of belonging to more 
than one category.  An outcome of this work is a system of statistical 
weights that can be used to compute global properties 
of galaxy samples.  We apply the MEx diagram to 
2,812 galaxies at $0.3<z<1$ in the Great Observatories Origins Deep 
Survey North and Extended Groth Strip fields, and compare it 
to an independent X-ray classification scheme.  
We identify Compton-thick AGN candidates with large X-ray absorption, 
which we infer from the luminosity ratio between hard X-ray emission and \oiiilam, 
a nearly isotropic tracer of AGN. X-ray stacking of sources that were not 
detected individually supports the validity of the MEx diagram and 
yields a very flat spectral slope for the Compton-thick candidates 
($\Gamma\approx$\,0.4, unambiguously indicating absorbed AGN).
We present evidence that composite galaxies, which are difficult 
to identify with alternative high-redshift diagrams, host 
the majority of the highly-absorbed AGN.  Our findings suggest that 
the interstellar medium of the host galaxy provides significant
absorption in addition to the torus invoked in AGN unified models.

\end{abstract}


\section{Introduction}

Most if not all galaxies contain a supermassive black hole (SMBH) in 
their center \citep{ric98}, which follow
fundamental relations to their host galaxies \citep{mago98,fer00}. 
Furthermore, the accretion history of supermassive black holes follow a similar, 
scaled-down, trend as the cosmic star formation history \citep{bar01}.
Taken together, these observations strongly suggest that black hole accretion and 
star formation may be linked phenomena.  In particular, active galactic nuclei 
(AGNs) have been invoked in galaxy evolution models as a means to control the rate 
of star formation in galaxies via feedback \citep[e.g.,][]{cro06,nar08}.  
While some observational evidence is provided by, e.g., high-velocity outflows in 
post-starburst galaxies \citep{tre07}, the full picture on the interplay between AGN 
and their host galaxies remains unclear.

A complete understanding of galaxy evolution requires the study of both galaxy stellar 
content and nuclear activity.  However, differentiating the powering source -- 
star formation versus AGN -- poses an appreciable challenge.
While there are several independent tracers of AGN, all suffer from limitations.
Because the caveats of one tracer can be overcome by the strength of another, 
it has become clear that multi-wavelength methods are required to assess 
the ubiquity of AGNs.
For example, X-ray emission has been used extensively to uncover and study large 
populations of AGNs \citep{bad95,boy93,bra01}.
However, even the most sensitive X-ray surveys still miss heavily absorbed systems.
These absorbed AGNs were inferred by the unresolved portion of the cosmic X-ray background, 
which shows a flat X-ray spectral slope highly suggestive of X-ray absorption 
\citep{com95,mus00,ale03,bau04,tre05}.  

This scenario is supported by the existence of Compton-thick systems in the 
nearby universe that would not be detected in the most sensitive X-ray 
surveys if they were at higher redshift, e.g., NGC~1068. This galaxy was 
also shown to have X-ray absorbers on a scale of the order of $\sim$1~pc 
\citep{gua00}.

Even if X-ray photons are absorbed or scattered by material on a compact 
scale such as the torus component described in the {\it unified model} 
\citep{ant93}, emission originating from larger scales may reach the 
observer regardless of the line of sight.  Such isotropic tracers include 
emission lines from the narrow line regions, which are exterior to the torus.  
Commonly used lines include \oiiilam\ in the optical \citep{bpt,bus88} 
and \oivlam\ or \nevmir\ in the infrared regime \citep{stu02,arm04,arm07,dia09,rig09,lama10}.
Another such tracer is mid-infrared continuum emission originating from hot dust 
heated by the AGN \citep[e.g.,][]{lac04,ste05,don07}.  This method works especially well
for intrinsically luminous AGN that are deeply enshrouded, but does not select
the less-luminous systems which may still be absorbed.  Mid-infrared aromatic features can 
also be used to diagnose the powering source in galaxies \citep{gen98,lut98,pop08}.

On the one hand, some of the distinctions between the classes of AGNs and 
the selection methods listed above are thought to arise from their orientation 
with respect to the observer's line of sight.
On the other hand, some AGN classes seem to comprise physically distinct 
phases of activity (i.e., low versus high accretion rate, beginning or 
end of an active phase).  The latter are more interesting from the point-of-view 
of galaxy evolution as they may provide a handle on the importance of AGN 
phases, their duty cycle, and the interplay between AGNs and their host galaxies.

One example of physically-distinct AGN phases are Seyferts and low-ionization 
nuclear emission line regions (LINERs).  These two categories emerged from 
optical spectroscopy studies where emission lines with different excitation 
properties were used as probes of the radiation 
exciting the interstellar gas \citep{sey43,hec80}.
In addition to exhibiting specific spectral signatures, Seyfert and LINER 
nuclei were found to reside in distinct host galaxies compared to star-forming 
galaxies and also relative to one another.  \citet{kau03c} showed that 
galaxies hosting an AGN tend to have a larger stellar mass compared to 
star-forming galaxies that lack optical AGN signatures.  Galaxies with a 
Seyfert nucleus (or Seyferts for short) often have a young or intermediate-age 
component in their stellar population whereas galaxies with LINER emission 
have, on average, an older stellar population as well as a larger stellar 
mass than Seyferts \citep[e.g.][]{kew06}.
Putting this evidence together with the observations that AGNs seem to 
follow a decreasing sequence in accretion rate from Seyferts to LINERs to 
composite galaxies \citep{ho08} suggests an evolutionary picture where 
LINERs may be older, dying, AGNs relative to Seyferts.

Whether this emerging picture is supported by higher-redshift observations 
is unclear.  At larger distances, it is generally more difficult to get a 
complete census of galaxies with AGNs, let alone to classify them in different 
AGN categories/phases.  Gathering complete samples of galaxies for which 
we know the powering source, and whether the central black hole is actively 
accreting, is especially challenging at $z > 0.4$.  Beyond that redshift,
optical emission lines needed for AGN classification such as \ha\ and \niilam\ 
are shifted into the near-infrared, preventing the application of well-calibrated, 
traditional diagnostics \citep{bpt,vei87,kew01,kau03a,kew06,sta06}.

In this paper, we present the Mass-Excitation (MEx) diagnostic diagram, based 
in part on optical nebular lines that can readily be observed out to $z \sim 1$.  
Following a similar method as \citet{wei07}, who replaced the \niilam/\ha\ 
line flux ratio used in the BPT diagram \citep{bpt} with absolute $H$-band 
magnitude, we adopt stellar mass as a substitute for \niilam/\ha.
We will show that a better census of AGN can be obtained by finding both 
intrinsically weak AGNs as well as absorbed systems that are undetected in 
X-ray observations. Our classification scheme relies on a novel probabilistic 
approach and allows us to split the galaxies into 
the following categories: purely star-forming galaxies, Seyfert 2s, 
LINERs or composite systems (i.e., with both star-formation and AGN). 
Galaxies near the empirical boundaries on traditional diagnostic diagrams have a 
less certain classification and are thus assigned a non-zero probability of belonging to more 
than one category.  As a result, the MEx diagnostic also outputs statistical 
weights that can be utilized to compute global properties (e.g., stellar mass, 
metallicity, etc.) in statistical samples of galaxies belonging 
to any of the categories listed above.

Using a sample of low-redshift galaxies described in \S\ref{sec:sdssSample},
we calibrate our diagnostic in \S\ref{sec:bpt}.  We briefly analyze 
the occurrence of low-ionization nuclear emission-line regions (LINERs) in 
\S\ref{sec:liner} before introducing a novel approach to galaxy spectral 
classification based on the probability of each spectral class (e.g., 
star-forming or AGN, \S\ref{sec:prob}).  We describe our $0.3 < z < 1$ 
galaxy sample in \S\ref{sec:sample}, and the application of our new 
Mass-Excitation (MEx) diagram in \S\ref{sec:hiz}.  We find an excellent 
agreement between the MEx diagram and the X-ray classification (\S\ref{sec:MEx_X}).  
We examine the different optical and X-ray classes more closely with an 
X-ray stacking analysis (\S\ref{sec:Xstack}).  Combining hard X-ray emission 
and optical emission lines allow us to probe the X-ray absorption leading 
to the discovery of Compton-thick AGNs among our intermediate-redshift sample 
\S\ref{sec:AGNpower}. We compare the MEx diagram with alternative AGN diagnostic 
diagrams in \S\ref{sec:compare} and we describe how the new method contributes 
to achieving a more complete census of AGNs (\S\ref{sec:census}). We discuss the fraction 
of AGNs that suffer from X-ray absorption (\S\ref{sec:absfrac}) and possible evolution effects on 
emission-line AGN diagnostics in \S\ref{sec:evol} before summarizing our 
main conclusions in \S\ref{summ}. 
We assume a flat cosmology with $\Omega_m = 0.3$, $\Omega_{\Lambda} = 0.7$, and 
$h = 0.7$ throughout.

\section{Low-Redshift Galaxy Sample}\label{sec:sdssSample}

Our low-redshift calibration sample comes from the Sloan Digital Sky Survey 
\citep[SDSS;][]{yor00}.  The limiting magnitude of the spectroscopic sample 
is $r < 17.7$.  Our analysis is based on data products from Data Release 4 
\citep{ade06}, namely the value-added galaxy 
catalogs\footnote{http://www.mpa-garching.mpg.de/SDSS/DR4/} from the Max-Planck 
Institute for Astronomy (Garching) and John Hopkins University.  In these 
catalogs, the stellar masses are calculated as described in \citet{kau03a} 
assuming a \citet{kro01} initial mass function (IMF), while emission line 
measurements follow the procedures from \citet{tre04}.

In order to avoid strong aperture bias due to SDSS fiber size, we constrain 
the redshift range to $0.05 < z < 0.2$.  This lower limit corresponds to a 
minimum covering fraction $\sim30\%$. \citet{kew08} found that using a covering 
fraction down to 20\% ($z\sim0.04$) is not sufficient to obtain global emission 
line properties for massive galaxies (with $M_{\star}>10^{10}~M_{\sun}$).  That 
is why we adopt a stricter requirement here.

Imposing a signal-to-noise ratio cut S/N$>3$ on all emission lines to be used 
in the diagnostic diagrams (\oiilam, \hb, \oiiilam, \ha, \niilam\ and \siilam; 
\S\ref{sec:bpt}), we obtain a sample of 110,205 emission-line galaxies.

\section{Emission-line Diagnostic Diagrams}

Here we aim to create a modified version of the BPT\footnote{Named after the 
last names of the three authors who introduced it: Baldwin-Phillips-Terlevich \citep{bpt}} 
diagram, involving \niilam/\ha\ and \oiiilam/\hb, by replacing the ratio of the 
redder emission lines (\niilam/\ha) because they shift into the near-infrared at 
$z > 0.4$.  To that purpose, 
we introduce the Mass-Excitation (MEx) diagram below, which we will subsequently 
apply to intermediate redshift galaxies ($0.3 < z < 1$) in \S\ref{sec:hiz}.

\subsection{Calibration Using $z \sim 0.1$ SDSS Galaxies}\label{sec:bpt}

\begin{figure*}
\epsscale{0.8} \plotone{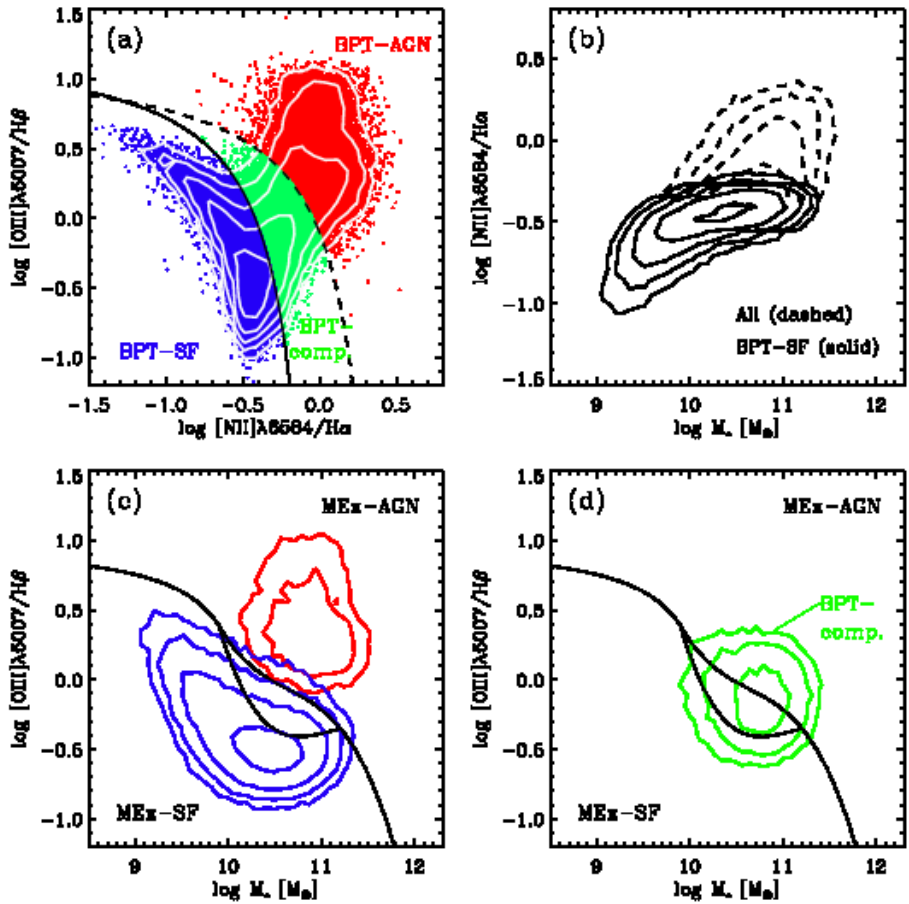}
\caption{Emission line diagnostic diagrams: (a) BPT diagram: purely star-forming 
  galaxies (SF, shown in blue) form a sequence below the solid line adapted from 
  \citet{kau03a} whereas galaxies hosting an AGN (red) tend to occupy the area 
  above and to the right of the dashed curve \citep{kew01}.  Galaxies located between
  the two curves are composites, i.e., having mixed SF and AGN contributions 
  (BPT-comp., green).  (b) Stellar mass as a function of the line ratio \niilam/\ha.  
  For star-forming galaxies the trend is analogous to the $M_{\star}-Z$ relation 
  (solid contours).  For the total sample including composites and AGNs, there is a 
  stronger increase in \nii/\ha\ compared to stellar mass but AGNs reside in 
  galaxies with both a high value of \nii/\ha\ and a high stellar mass. These two 
  features indicate that stellar mass is a viable substitute when \niilam\ or 
  \ha\ are not available. 
  The modified diagnostic diagram is shown in (c) for the galaxies classified as 
  SF or AGN on the BPT diagram (blue and red contours, respectively) and in (d) for 
  those classified as composites (green contours).  
  The MEx-intermediate region, located between the two empirical curves on the MEx diagram, contains a mix of 
  composites (BPT-comp.) and star-forming (BPT-SF) galaxies.  In all panels, the 
  contours indicate the density of points (in bins of 0.075 dex $\times$ 0.075 dex) 
  and are logarithmic (0.5~dex apart, with the outermost contour set to 10 galaxies per bin).
  (A color version of this figure is available in the online journal.)
   }\label{fig:BPT}
\end{figure*}

In this section, we introduce and calibrate the Mass-Excitation diagram 
and compare it to one of the original BPT diagrams.  
The emission line ratios used in the best-known version 
of this diagram (\niilam/\ha\ and \oiiilam/\hb) probe a combination of the 
ionization parameter and the gas-phase metal abundance within galaxies.
As shown in Figure~\ref{fig:BPT}(a), SDSS galaxies form a well-defined 
excitation sequence on the lower-left of the BPT diagram \citep[below the 
semi-empirical dividing curve from][solid line]{kau03a}, while the galaxies 
containing AGN form a plume extending to the top right part of the diagram.  
The higher ionization parameter and/or harder ionizing radiation that 
occur only in the presence of AGN cause the line ratios to lie 
above and to the right of the 
maximum starburst curve (dashed line) developed by \citet{kew01}.  
Galaxies that are located between both curves are believed to host a
mixture of star-formation and AGN and are sometimes called 
composites.  We adopt this nomenclature in the remainder of this work.

While the BPT diagram shown in Figure~\ref{fig:BPT}(a) is used 
extensively to identify the source of ionization in galaxies, the 
\nii\ and \ha\ emission lines become unavailable in optical spectra 
at $z>0.4$.  
What would be a good substitute for \nii/\ha? The \nii/\ha\ line 
ratio provides an indication of the gas-phase metallicity in star-forming 
galaxies \citep{kew08}. The empirical mass-metallicity relation 
\citep[e.g.,][]{tre04,sav05} suggests a physical connection between 
that line ratio and the stellar mass, as displayed in Figure~\ref{fig:BPT}(b). 
While the $M_{\star}-Z$ relation 
applies to star-forming galaxies without AGNs, there is another effect 
that makes stellar mass a good choice.  The \nii/\ha\ ratio saturates at 
high values for normal star-forming galaxies and only more extreme 
conditions such as those encountered in the presence of an AGN can yield 
larger values \citep{kew06,sta06}.  Because AGNs tend to be found in hosts 
with high stellar mass \citep{kau03c}, these systems have both larger 
\nii/\ha\ and $M_{\star}$ values.  This feature puts them in a location of 
the parameter space of the modified diagram that is analogous to their 
original location on the BPT diagram with respect to purely star-forming 
galaxies (i.e., higher and to the right).  
Consequently, the new AGN diagnostic is obtained by substituting stellar mass 
for the redder emission line ratio.  

Indeed, we find that the BPT-SF and BPT-AGN classes are well separated 
on the new Mass-Excitation (MEx) diagram (Figure~\ref{fig:BPT}(c)).  
We define two empirical dividing lines that maximize the separation between 
galaxy classes, especially between the BPT-AGN class (above and to the 
right of the lines) and the purely star-forming BPT-SF class (below 
and to the left of the dividing lines).
We note that the location of the BPT-composite galaxies on the MEx diagram 
overlap with galaxies belonging to the other classes as shown in 
Figure~\ref{fig:BPT}(d). Their locus peaks in the region between the 
two empirical curves, which we dub the MEx-intermediate region.  We 
note that 48\% of the galaxies in this region are BPT-composites.  

The number of galaxies of each BPT-class are reported in Table~\ref{tab:diag} 
for the three regions of the MEx diagram, and the main empirical dividion 
(top curve) is defined as follows:  
\begin{equation}\label{eq:mexline}
y = \left\{ \begin{array}{ll}
      0.37/(x - 10.5) + 1. & \mbox{if x\ensuremath{\leq}9.9} \\
      a_0 + a_1 x + a_2 x^2 + a_3 x^3 & \mbox{otherwise,}
      \end{array}
     \right. 
\end{equation}
where $y \equiv$ log$(\oiiilam/\hb)$ and $x \equiv$ log$(M_{\star})$. 
The coefficients are the following: $\{a_0, a_1, a_2, a_3\} = 
\{594.753, -167.074, 15.6748, -0.491215\}$.
Similarly, the lower curve defined as:
\begin{equation}\label{eq:mexlower}
y = 800.492 - 217.328x + 19.6431x^2 - 0.591349x^3,
\end{equation}
and is used over the range $9.9<x<11.2$.

\begin{figure}
 \epsscale{0.85} \plotone{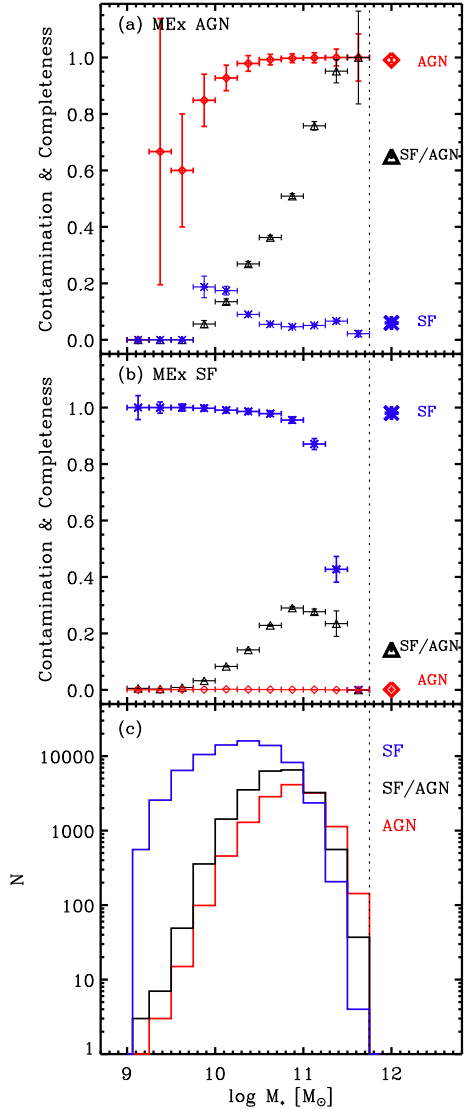}
\caption{
   Completeness and contamination rates in bins of stellar mass (0.25~dex bins).
   {\it Top:} For the MEx-AGN selection, the AGN completeness is defined as the 
   percentage of BPT-AGNs (above the Kewley line in the BPT diagram) 
   that are above the MEx diagram dividing line (red diamonds).  The definition 
   is similar for the BPT-composite galaxies (black triangles).
   The contamination rate corresponds to the percentage of galaxies in the AGN 
   side of the MEx diagram that would be purely star-forming galaxies according 
   to the BPT diagram (blue asterisks).  The overall values for the entire stellar 
   mass range are shown on the right hand side of the figure.
   {\it Middle} For the MEx-SF selection, the completeness is defined as the 
   percentage of BPT-SF galaxies (below the Kauffmann line) that are 
   correctly identified in the star-forming side.  The contamination rates are 
   computed separately for the BPT-AGNs (red diamonds) and BPT composites (black 
   triangles) and correspond to percentage of the number of galaxies in the star-forming
   side of the MEx diagram.  Error bars are Poissonian. {\it Bottom} Distribution of  
   stellar masses for galaxies that are classified as star-forming (blue), AGN (red), or
   composite (black). 
   (A color version of this figure is available in the online journal.)
}\label{fig:complet}
\end{figure}

Using the BPT classes as our reference, we compute the completeness and 
contamination fractions of the MEx selection for both the AGN and SF classes, 
separated using Eq.~\ref{eq:mexline}.  The fractions are displayed in 
Figure~\ref{fig:complet}.  For the MEx-AGN class, we find that the completeness 
is very high in terms of BPT-AGNs (i.e., above the \citet{kew01} line), 
reaching close to 100\% at high stellar mass (see the red diamonds in the 
top panel of Figure~\ref{fig:complet}).  The fraction of BPT-composite galaxies 
that are selected in the MEx-AGN side rises more slowly with stellar mass: 
from $\sim$20\% at $M_{\star} < 10^{10}~M_{\sun}$ to 95\% at $M_{\star} > 10^{11.5}~M_{\sun}$ 
(black triangles).  In the top panel, the contamination fraction is defined 
as the fraction of BPT-SF galaxies (below the \citet{kau03c} line) in the 
MEx-AGN side.  The contamination fraction peaks at 10-20\% for galaxies with 
$M_{\star} \sim 10^{10}~M_{\sun}$, but it mostly stays well below 10\%, with an 
overall fraction around 6\% (blue asterisks).  Global completeness and 
contamination values, calculated for the entire SDSS sample (all stellar masses), 
are shown with the larger plotting symbols on the right hand side of the figure.


Similarly, the completeness for the MEx-SF selection is defined as the fraction of 
the BPT-SF galaxies that are correctly identified.  The completeness is close to 
100\% (blue asterisks in Figure~\ref{fig:complet}[b]) over a wide range of 
stellar masses with a drop off at $M_{\star}> 10^{11}~M_{\sun}$. We note that at 
such high masses, there are very few purely star-forming galaxies ($\sim$200 and $<$10 in 
the last two bins) whereas there are of the order of 10$^{4}$ galaxies per 
bin at lower mass (Figure~\ref{fig:complet}[c]). The contamination with the 
BPT-AGN class is extremely low (red diamonds), on the order of 0.3\%.  However, 
the contamination fraction for BPT-composite goes up to $\sim$30\% at 
$M_{\star} > 10^{11}~M_{\sun}$.  These galaxies lie mostly in the intermediate 
region of the MEx diagram.  Whether they should be included with the star-forming 
sample or rejected will depend on the specific science goal.

We report the numbers for the MEx-intermediate class (i.e., between the lines defined 
by Eq.~\ref{eq:mexline} and \ref{eq:mexlower}) separately in Table~\ref{tab:diag}.  
Even though the AGN contamination of the MEx-SF class appears to be more substantial 
at high stellar mass, the effect is exaggerated by the small number of purely 
star-forming galaxies with $M_{\star}> 10^{11}~M_{\sun}$.  In fact, the AGN completeness 
peaks at the same mass range where the MEx-SF class appears the most contaminated, 
indicating that overall, the majority of the most massive galaxies harbor an AGN. 
Conversely, very few AGNs reside in low-$M_{\star}$ hosts. 

The scarcity of BPT-AGNs in host galaxies with $M_{\star} <10^{10}~M_{\sun}$ 
was previously noted by \citet{kau03c}.  These authors found that adding 
the emission from low-luminosity AGNs (with $10^5 < L_{\oiiilam} < 10^6~L_{\sun}$) 
to low-$M_{\star}$ star-forming galaxies would significantly alter their 
line ratios and move the corresponding points into the composite or AGN 
regions of the BPT diagram (in 93\% of the cases with low-luminosity AGNs, 
and $>$99\% for high-luminosity AGNs, i.e., with $L_{\oiiilam} > 10^7~L_{\sun}$).

\begin{figure*}[t]
\epsscale{1.} \plotone{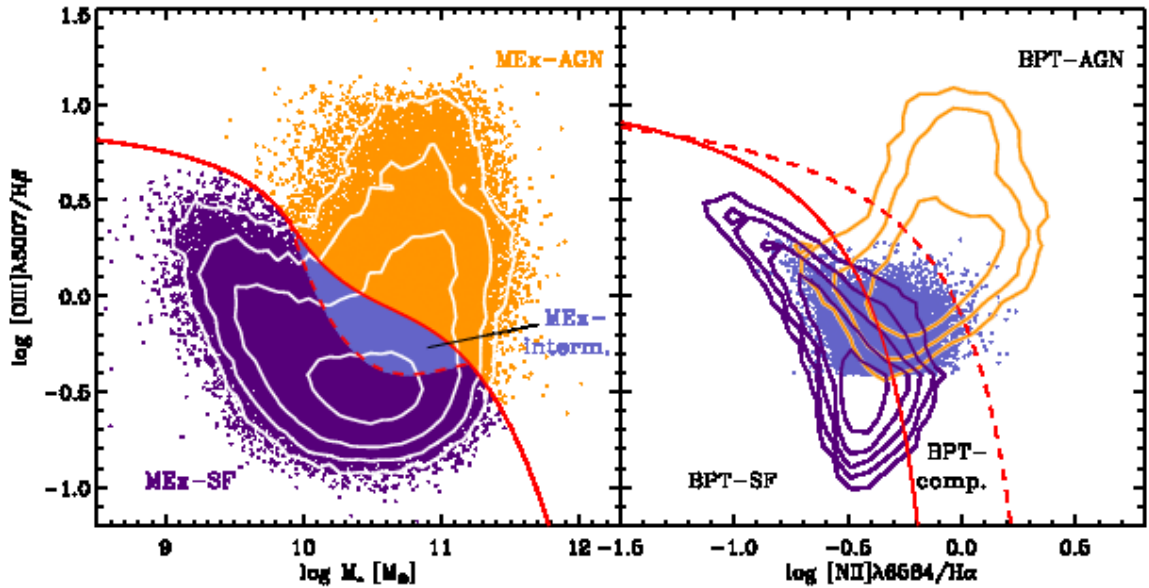}
\caption{(left) Distribution of SDSS galaxies on the MEx diagram with our 
   empirical divisions. Galaxies above the lines as classified as MEx-AGN (orange) 
   whereas the galaxies below the lines are classified as star-forming (MEx-SF, 
   in violet) and galaxies between the lines are MEx-intermediate (MEx-interm., in 
   light purple).  Using the same color-coding, we show these galaxies on the standard 
   BPT diagram (right).  Here, galaxies that are in the MEx-intermediate region 
   are shown with dots while the MEx-SF and MEx-AGN subsamples are overplotted with 
   contours (violet and orange, respectively). The contours indicate the density 
   of points (in bins of 0.075 dex $\times$ 0.075 dex) and are logarithmic (0.5~dex 
   apart, with the outermost contour set to 10 galaxies per bin).
   (A color version of this figure is available in the online journal.)
\label{fig:BPT2}}
\end{figure*}

So far, we have used the BPT spectral types (SF, AGN, composites) as references 
to quantify the completeness and contamination rates of the MEx selection.  
Now, we turn the situation around and we show a graphical comparison of the 
MEx diagram selection (eq. 1) mapped back onto the usual BPT diagram.
Figure~\ref{fig:BPT2}(a) shows the classification on the MEx diagram with 
MEx-AGN and MEx-SF galaxies colored in orange and purple, respectively.  
The contours show the density of points on a logarithmic scale.  As can be seen 
in panel (b), our new AGN selection (orange contours) picks all the AGNs 
from the BPT diagram (above the dashed line in panel (b)) as well as a 
fair number of BPT-composites (mostly at higher \oiiilam/\hb\ ratio), whereas 
our new MEx-SF selection (purple contours) captures the BPT star-forming sequence 
extremely well, with an extension into the BPT-composite region between the 
two dividing lines, especially in the region close to the \citet{kau03c} solid line.
Galaxies in the MEx-intermediate region (light purple dots) are distributed 
over the BPT-comp. and BPT-SF regions.  

\subsection{A closer look at LINERs}\label{sec:liner}

The nature of LINERs remains controversial.  Historically, they have been identified 
by the unusual strength of their narrow optical emission lines with low-ionization 
potential \citep{hec80}. While most studies consider that LINERs are accretion-powered 
\citep{ho99,kew06}, there are also claims that, in some cases, the powering source could 
be entirely stellar \citep[e.g., from post AGB stars and/or white dwarfs, ][]{bin94,sta08} 
or a combination of processes including shocks \citep{hec80}. 
\citet{era10} argue that AGN activity does not provide enough energy to produce the LINER 
emission in half of their X-ray-selected sample of 35 LINERs.  Such a deficit had been suggested 
previously \citep[e.g.,][]{ho93}, and could be compensated by either obscuration of UV 
photons from the AGN, or by contributions from alternative sources such as post-AGB stars.

Chandra X-ray observations have been used to look for X-ray cores within LINERs.  
The detection fraction is typically high, ranging from 50\% to 70\% 
\citep[see the review by][and references therein]{ho08}.  Employing {\it Spitzer} 
spectroscopy, \citet{dud09} searched for high-ionization lines associated with AGN 
activity ([Ne V] 14 and 24\,$\mu$m).  They find that 39\% of their sample of 67 LINERs 
have such detections and that many of these cases lack AGN signatures at optical and 
X-ray wavelengths.  They also show that the optical identification  (from, e.g., broad 
\ha\ lines) is more subject to fail at higher infrared luminosities, suggesting that 
some AGNs may be missed due to dust obscuration.  Their AGN fraction in LINERs goes up 
to 74\% after combining diagnostics in all three wavelength ranges (X-ray, optical, mid-IR).

\begin{figure*}
\epsscale{0.85} \plotone{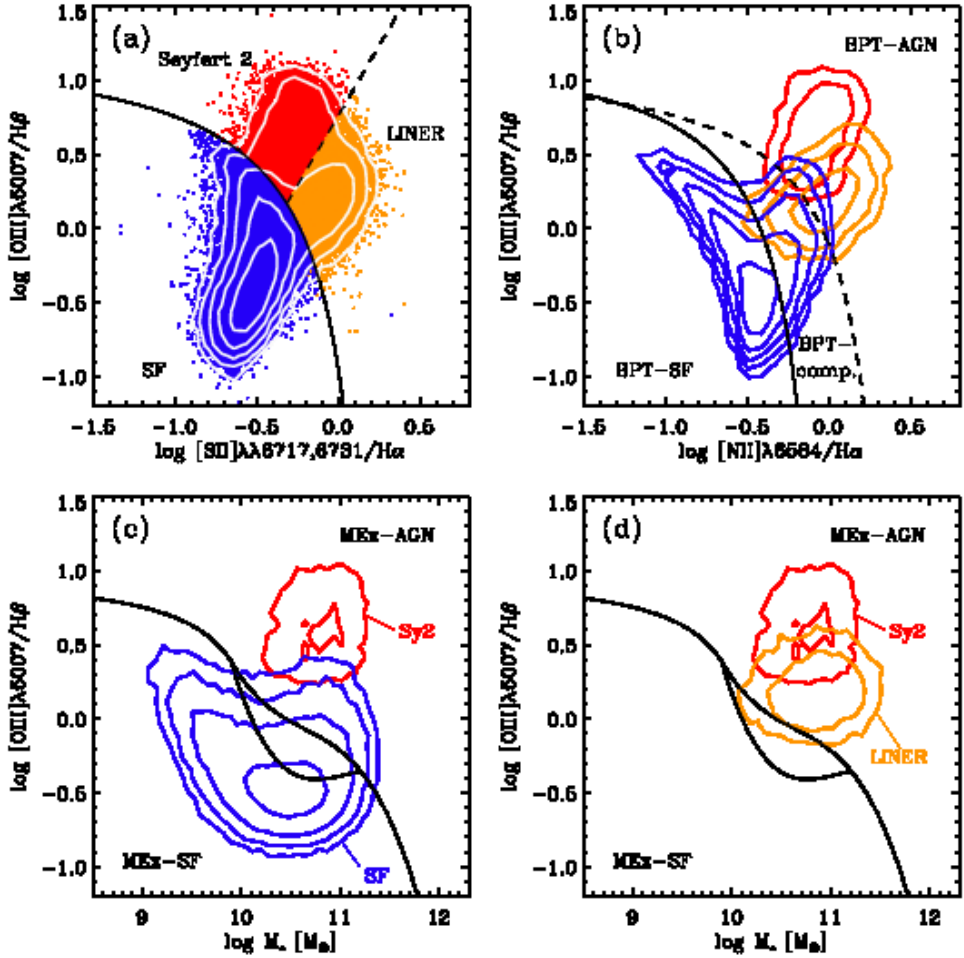}
\caption{(a) The \sii\ diagram.  SDSS galaxies are color-coded according to their classification 
   as SF (blue), Seyfert 2 (red) or LINER (orange).  The dividing lines are taken from 
   \citet{kew06}.  All three classes are shown on the BPT diagram (b) with the same color scheme.
   The \sii\ diagram is not very sensitive to composite galaxies (BPT-comp.) as most of them are 
   classified as SF (blue contours).  Similarly, the three \sii\ diagram classes are shown
   on the MEx diagram (panel (c) for the SF and Seyfert 2 galaxies, and panel (d) for the LINERs 
   relative to the Seyfert 2's).
   The upper empirical line on the MEx diagram selects all Sy2's and LINERs on the 
   AGN side (above the lines), with some overlap between the two classes. In all panels, 
   contours indicate the density of points 
   (in bins of 0.075 dex $\times$ 0.075 dex) and are logarithmic (0.5~dex apart, with the 
   outermost contour set to 10 galaxies per bin).
   (A color version of this figure is available in the online journal.)
\label{fig:liner}}
\end{figure*}

The BPT diagram that includes \oiiilam/\hb\ against \siilam/\ha, hereafter the 
\sii\ diagram, can be used to tell apart the LINERs from the Seyfert 2 
(Sy2) galaxies \citep[e.g.,][]{kew06}.  We apply this diagram to our SDSS 
sample in Figure~\ref{fig:liner}.  Panel (a) shows the \sii\ diagram 
with dividing lines between star-forming, Seyfert, and 
LINER populations as labeled. Seyfert 2 (in red) and LINER (in orange) galaxies 
as defined in the \sii\ diagram overlap slightly on the \nii\ (panel b) and 
MEx (panels c,d) diagrams.  BPT-composites (between the lines on the BPT \nii\ 
diagram) fall closer to the LINER than to the Sy2 distribution.

Composite galaxies stand out more in their \nii/\ha\ than in their \sii/\ha\ ratio.  
Consequently, the star-forming selection (blue contours) extends into the composite 
and AGN regions of the BPT and MEx diagrams [Figure~\ref{fig:liner}(b) and (c)].  
As mentioned previously, the \sii/\ha\ ratio has the advantage of splitting the LINERs 
from the Seyfert AGNs [dashed line from \citet{kew06}].  This allows us to see that the 
MEx diagram selects both of these types of AGNs, and that the SDSS sample used here contains 
more LINERs than Seyfert 2's.

In the remainder of our analysis, we will use a scheme that combines 
the most trusted features of each diagram (i.e., the \nii\ and the \sii\ 
diagrams), which we introduce in the next section.

\subsection{A Probabilistic Approach to Galaxy Classification}\label{sec:prob}

As was shown previously, some galaxy spectral classes overlap on the MEx diagram 
[Fig.~\ref{fig:BPT}(b,c); \ref{fig:BPT2} and \ref{fig:liner}(c,d)].  
Namely the MEx-intermediate region contains BPT-SF and BPT-composites, and there 
is also overlap between the BPT-LINERs and other AGN subclasses.  
To better assess the classification in such cases we present a scheme based on the 
probability of each spectral class given a galaxy's position on the MEx diagram.
This approach is useful to discriminate between star-forming galaxies, composite 
galaxies, LINERs and Seyfert 2's.  

In order to quantify the probabilities, we use a 
low-redshift\footnote{0.05$<z<$0.1; The upper bound of the redshift 
range is chosen to remain complete in all categories including LINERs,
whose detection rate decreases beyond $z \sim 0.1$ \citep{kew06}.}
SDSS sample as a calibration set. All of the BPT diagnostic emission 
lines are available for these galaxies and we can place them on the 
MEx diagram with prior knowledge of their source of ionization.  
The source of ionization is assigned according to a hybrid classification 
based on both the \nii\ and \sii\ BPT diagrams with the following rules:
\begin{enumerate}

\item[1.] Star-Forming (SF): galaxies below the \citet{kau03a} line on the \nii\
   diagram irrespectively of their class on the \sii\ diagram.

\item[2.] LINER: galaxies that are classified as AGN in the \nii\ diagram 
   \citep[above the line from][]{kew01} or in the \sii\ diagram and that are below
   and to the right of the Seyfert/LINER dividing line on the \sii\ diagram (Eq.~13
   of \citet{kew06}).

\item[3.] Seyfert 2 (Sy2): galaxies that are classified as AGN in the \nii\ diagram 
   \citep[above the line from][]{kew01} or in the \sii\ diagram and that are above 
   and to the left of the Seyfert/LINER dividing line on the \sii\ diagram (Eq.~13
   of \citet{kew06}).

\item[4.] Composite (comp): composite galaxies from the \nii\ diagram
   (between the \citep{kau03a} and \citep{kew01} curves) that were not included in 
   the LINER and Sy2 classes described above.

\end{enumerate}

The resulting spectroscopically-classified sample is the 
{\it SDSS prior sample}, which we assume is representative of 
galaxies out to $z=1$ (see \S\ref{sec:evol} for a discussion of 
possible evolutionary effects).

Relative to the previous diagrams, the combined classification method 
yields a slightly greater number of AGNs than each of the \nii\ and 
\sii\ diagram taken separately (because we use the union of the two 
AGN subsamples).  Correspondingly, the composite class is slightly 
less numerous because some composite galaxies from the \nii\ diagram 
are selected as AGN based on the \sii\ diagram (mostly 
in the LINER region).  Another distinction is that the LINERs from the \sii\ diagram that 
extend into the star-forming side of the \nii\ diagram are now in the star-forming category.
The latter category is identical to the original \nii\ star-forming selection.

We now have a spectral classification scheme that exploits the strengths of both the \nii\
and \sii\ diagrams, and contains four mutually exclusive categories: star-forming (SF), 
composite (comp), LINER and Seyfert 2.  We employ the distribution of the SDSS galaxies 
on the MEx diagram with {\it a priori} knowledge of their BPT classification to compute 
the fraction of galaxies of each category across the MEx plane.

For galaxies at higher redshift for which only \oiiilam/\hb\ and stellar mass 
($M_{\star}$) are available, we use the rectangular region on the MEx diagram 
defined by the one-sigma uncertainties on these two measurements.  We compute 
the number of {\it SDSS prior} galaxies in each category (star-forming, composite, 
LINER, and Seyfert~2) and we normalize by the total number of {\it SDSS prior} galaxies 
within the rectangular box.  The fractions are converted to percentages.  Given that the four 
classes described above are mutually exclusive, the sum $P(SF)+P(comp)+P(LINER)+P(Sy2)=100\%$.  
For example, if a region surrounding given values of  \oiiilam/\hb\ and 
$M_{\star}$ (defined by the one-sigma error bars) contains 20,000 SF galaxies; 10,000 
composites; 10,000 LINERs and no Seyfert 2s in the SDSS prior sample, the 
assigned probabilities would be $P(SF)=50\%$, $P(comp)=25\%$, $P(LINER)=25\%$, 
and $P(Sy2)=0\%$.  Thus, this probabilistic AGN classification scheme has a 
built-in uncertainty.  This is a useful feature compared to alternative diagrams 
where there is often no knowledge of the reliability of a certain classification.  
With this new approach, we know whether a given galaxy is near a dividing line 
or whether it is located far into the AGN or star-forming locus.  
In this paper, we assume that composites, LINERs, and Seyfert 2's all host an 
active nucleus and often use: $P(AGN) = 1-P(SF)$ (equivalent to 
$P(AGN)=P(comp)+P(LINER)+P(Sy2)$).

\begin{figure*}
\epsscale{1.} \plotone{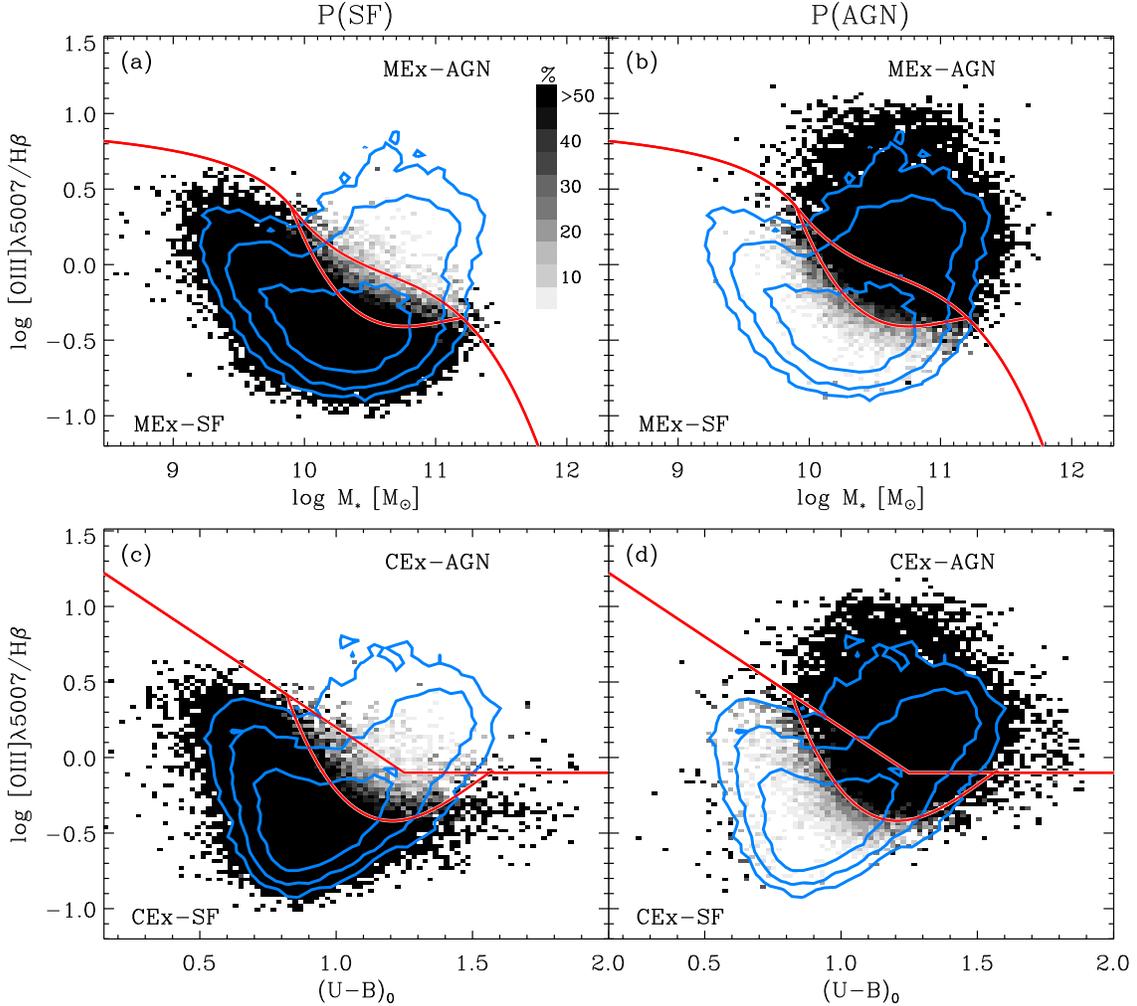}
\caption{
   Percentages of galaxies classified as star-forming or AGN (all sub-categories) as a function 
   of location on the MEx and CEx diagrams (top and bottom row, respectively).
   The left-hand side shows the number of SF galaxies divided by total number galaxies, or $P(SF)$, 
   in each bin (0.04~dex $\times$ 0.04~dex) for the MEx (a) and CEx (c) diagrams. 
   The AGN percentages $P(AGN)$ are displayed on the right hand side, where panel (b) is the MEx diagram
   and panel (d) the CEx diagram developed by Y11.  ($U-B$)$_0$ denotes the rest-frame
   $U-B$ color expressed in AB magnitudes.  In all panels, the upper lines 
   mark the main division between AGN and star-forming galaxies, whereas the lower lines 
   approximately correspond to a transition from $P(AGN)<30\%$ (below) to $P(AGN)>30\%$ (above).  
   Galaxies between both sets of lines are a mixed SF/composite population, 
   which we call MEx-intermediate galaxies (or CEx-intermediate in panels c,d).
   The bivariate distribution of the SDSS sample combining all galaxy classes is shown 
   with logarithmic contours. 
   (A color version of this figure is available in the online journal.)
}\label{fig:prob}
\end{figure*}

The empirical division on the MEx diagram introduced earlier (\S~\ref{sec:bpt}) traces well 
the observed transition between SDSS galaxies that host AGN activity of any category 
(composite, LINER, or Seyfert 2) and galaxies that are most likely star-forming 
[Figure~\ref{fig:prob}(a,b)].  Figure~\ref{fig:prob} shows the probabilities $P(SF)$ 
and $P(AGN)$ as a function of position on the MEx diagram.  
The lower dividing curve delineates the separation between a \emph{cleaner} star-forming 
galaxy sample (below) and the MEx region where $P(AGN)>30\%$ (above).  
The region between the two curves contains a mixed BPT-SF/BPT-composite population.  
We adopt the term MEx-intermediate to describe this region of the MEx diagram and 
the galaxies that are located within it.

An alternative diagnostic diagram, developed in parallel to the MEx diagram, involves the same 
emission-line ratio on the vertical axis but makes use of rest-frame $U-B$ color
rather than stellar mass \citep[][hereafter Y11]{yan11}.  We display the AGN and 
star-forming fractions on this color-excitation diagram from Y11 in a similar fashion as for the 
MEx diagram [Figure~\ref{fig:prob}[c,d)].  The original dividing lines are adapted
from Y11 (straight lines) and the definition is included here for completeness:
\begin{equation}
\log(\oiiilam/\hb) = max\{1.4-1.2(U-B), -0.1\}.
\end{equation}

In this paper, we add a curve on the diagram from Y11 that follows the transition where 
the AGN probability is $P(AGN)>30\%$.  The new region between this curve and the 
straight lines is analogous to the MEx-intermediate region of the MEx diagram and 
contains $>50\%$ of BPT-composite galaxies.  
The lower curve in Figure~\ref{fig:prob}(c,d) is defined by:
\begin{equation}
y = 12.3914 - 27.0954x + 18.5122x^2 - 4.02369x^3,
\end{equation}
where $y \equiv$ log$(\oiiilam/\hb)$ and $x \equiv$ ($U-B$)$_0$. As in Y11,
we calculate the rest-frame $U-B$ color, expressed in AB magnitudes, by using the 
{\it k-correct} v4\_1\_4 code from \citet{bla07}.

The bivariate distributions of galaxies on both the mass-excitation (MEx) and the color-excitation
(CEx) diagrams are examined in more detail in Appendix~\ref{app:class}.

The SDSS subsample used here has been implemented as a reference to provide 
probabilities of each class as a function of the location on the MEx (or CEx) diagram.
Given a stellar mass and \oiiilam/\hb\ ratio, our publicly available 
IDL\footnote{Interactive Data Language.} 
routines\footnote{http://monkey.as.arizona.edu/$\sim$sjuneau/AGN-Galaxy Classification.html} 
return a probability that the input galaxy belongs to the SF, composite, LINER, and 
Seyfert 2 categories.  
The MEx diagram is well motivated from the successes at separating and quantifying the 
overlap of BPT-classes for a large SDSS sample of galaxies.  
We will apply this new diagnostic to a sample of galaxies at 
$0.3 < z < 1$ in the following sections.

\section{Intermediate-Redshift Galaxy Sample}\label{sec:sample}

Our intermediate-redshift galaxy sample is based on observations from 
the Great Observatories Origins Deep 
Survey\footnote{http://www.stsci.edu/science/goods/} (GOODS) and the 
All-wavelength Extended Groth strip International 
Survey\footnote{http://aegis.ucolick.org/} (AEGIS).  Most of the analysis 
is performed on galaxies at $0.3 < z < 1$, although we extend the 
range to slightly lower values when we use redder emission lines 
(such as in \S\ref{sec:MEx_X}).

Optical spectra are drawn from the Team Keck Redshift 
Survey\footnote{http://tkserver.keck.hawaii.edu/tksurvey/} \citep[TKRS][]{wir04}
for galaxies in the GOODS-North (GOODS-N) field, and from the 
DEEP2 Galaxy Redshift Survey \citep[hereafter DEEP2;][]{dav03,dav07} 
for galaxies in the Extended Groth Strip (EGS) field.  These two 
spectroscopic surveys have similar limiting magnitudes of $R_{AB}=24.3$ 
and $24.1$, respectively.  The former lies within GOODS-N (10\arcmin 
$\times$ 16\arcmin) and includes redshifts for 1440 galaxies 
(1044 galaxies with confident redshifts $0.3 < z < 1$).  From the 
DEEP2 survey, we only use the pointing in the Extended Groth Strip 
(centered at 14 17, +52 30) covering 120\arcmin $\times$ 15\arcmin.
There are 6,588 DEEP2 galaxies with confident redshifts $0.3 < z < 1$ 
in this pointing.

Both sets of observations were obtained with the DEIMOS spectrograph 
\citep{fab03} at the Keck Observatory and reduced with the 
pipeline\footnote{http://astro.berkeley.edu/$\sim$cooper/deep/spec2d/} developed 
by the DEEP2 team at the University of California-Berkeley.
However, their spectral resolution and spectral range differ due 
to the use of different gratings (600 line~mm$^{-1}$ for TKRS and 
1200 line~mm$^{-1}$ for DEEP2).  The TKRS resolution is 
4\AA~full-width-half-maximum (FWHM) over the wavelength range 
5500$-$9800\AA, whereas DEEP2 spectra have a resolution of 2\AA~FWHM 
with a wavelength coverage of 6500$-$9100\AA.  These different 
spectral ranges mean that emission lines of interest are accessible 
over somewhat different redshift ranges for the two samples.  \hb\ and 
\oiiilam\ can be observed out to $z \sim$1 with TKRS spectra but only 
out to $z \sim$0.8 with DEEP2 spectra.  

A wealth of multiwavelength data is available in both GOODS-N and EGS.  
In this paper, we utilize ancillary {\it Chandra} X-ray data, {\it Spitzer} 
IRAC data, and ground-based optical imaging (\S\ref{sec:mass}). 
{\it Spitzer}/IRAC photometry is available in all four channels 
\citep[available through the {\it Spitzer} Science Center, also see][for EGS]{bar08}.  
In what follows, IRAC photometry is used to estimate the rest-frame 
$K$-band magnitude (\S\ref{sec:mass}).  We take advantage of the fact 
that the {\it Chandra} X-ray coverage is very deep: 2~Msec in GOODS-N \citep{ale03} 
and 200~ksec in EGS \citep{lai09,nan05}.  This provides us with an 
independent AGN/star-forming classification scheme (\S\ref{sec:hiz}).  
The sensitivity of the shallower data is adequate to ensure the detection 
of luminous X-ray sources ($L_{2-10{\rm keV}} > 10^{42}$~erg s$^{-1}$, a nominal 
luminosity threshold for AGN) out to $z \sim 1$.  Furthermore, we can 
detect fainter X-ray galaxies (including starbursts) at all redshifts 
in GOODS-N.

\subsection{Emission Line Fluxes}

Emission line fluxes from the TKRS and DEEP2 spectra were measured using 
automated IDL routines.  For each emission line, we fit the continuum with 
a linear relation over 30~\AA~windows on either side of the line with a 
biweighting scheme.  This technique efficiently rejects outliers and is 
thus robust against pixels with large errors (e.g., due to large sky 
subtraction residuals).  If the flux density at the line peak is greater 
than three times the standard deviation of the continuum, the line is fitted 
with a Gaussian function.  In such cases, we calculate emission line 
fluxes in two ways.  First, we integrate the resulting Gaussian fit over 
a wavelength range corresponding to $\pm 2.5\sigma$, where $\sigma$ is 
the Gaussian width parameter (= FWHM/2.35).  Secondly, we directly integrate 
the continuum-subtracted spectra over the same wavelength range.

In most cases, we use the fluxes obtained from integrating the data 
directly.  However, some potentially problematic cases were flagged 
for visual inspection.  Among 509 TKRS galaxies for which both the \hb\ 
and \oiiilam\ emission lines passed the signal-to-noise and quality criteria, 
61 were flagged as uncertain.  Similarly, 245 among 2,536 DEEP2 galaxies were 
flagged for visual inspection.  This generally occurs when the data are corrupted 
nearby the targeted emission line, or when the Gaussian fit is inadequate.  
These objects are examined on a case-by-case basis, and the flux measurement 
is adjusted accordingly.  Among a total of 306 galaxies that were examined, 
184 were validated whereas 122 were flagged as uncertain and not used in 
subsequent analysis.

We correct for underlying stellar absorption at \hb\ and \ha.  TKRS spectra with 
a median signal-to-noise ratio (S/N) per pixel greater than three were fit 
individually using \citet[hereafter BC03]{bru03} spectral synthesis models.  
Utilizing IDL {\it simplefit} routines (C. Tremonti, private communication), 
we fit the continuum of the galaxies with a linear combination of ten representative 
stellar population templates, leaving dust obscuration as a free parameter.  
We subtract the continuum to correct Balmer lines for underlying stellar absorption.  
The median corrections ($\pm$ half of $84^{th}-16^{th}$ percentile range), expressed 
in terms of equivalent widths (EWs), are $2.8 (\pm0.9)$~\AA~at \hb\ and $1.4 (\pm0.7)$~\AA 
at \ha.  These values were applied to spectra that were not fit individually due to 
low signal-to-noise ratio or uncertain spectrophotometry.  We note that applying the 
median correction to \hb\ (\ha) for spectra with an individual fit changes their line 
fluxes by 0.08 (0.03)~dex r.m.s.

In addition to using line ratios, we will use \oiiilam\ luminosities to quantify the 
strength of AGN activity (e.g., in \S\ref{sec:AGNpower}).  For that purpose, the slit 
loss corrections are obtained by calculating synthetic photometry from the spectra in 
the band nearest to the observed wavelength of \oiiilam (usually ACS $F775w$ or $F814w$ 
for GOODS-N and EGS observations, respectively) and comparing to the true observed photometry.  
The synthetic photometry is obtained by applying the filter curve and integrating over 
the wavelength range of interest in the observed band.  The ratio between the total 
flux from observed photometry and the synthetic photometry is used as a multiplicative 
correction factor.  The median slit loss correction is a factor of two.

\subsection{Stellar Masses}\label{sec:mass}

The stellar masses of our intermediate-redshift sample were calculated 
by fitting stellar population synthesis models to spectral energy distributions (SEDs)
measured by galaxy photometry.  The procedure is described in \citet{sal07}.  
For galaxies in EGS, the following photometric bandpasses are used: 
FUV, NUV (GALEX), $ugriz$ (CFHTLS), and $K$ (Palomar) \citep[see][]{sal09,gwy08,gwy11,bun06}.
For galaxies outside of the Canada-France-Hawaii Telescope Legacy Survey (CFHTLS) 
field-of-view, we use CFHT 12k $BRI$ photometry from \citet{coi04}.
For GOODS-N, the constraints are provided by the following 
photometry: $UBVRIz$ taken from \citet{cap04} and $JK$ obtained 
with the Flamingos camera on the Mayall 4~m NOAO telescope.
In GOODS-N, the SED fitting is performed for galaxies with $K<20.5$ (Vega)
\citep[as in][but extending to lower redshifts]{dad07}.

The calculations assumed a Chabrier IMF \citep{cha03}, and output a probability 
distribution function (PDF) for the stellar mass.  We assume the average of the 
PDF as the stellar mass, and estimate errors from the 2.5th and 97.5th percentiles 
[$=(97.5PL - 2.5PL)/3.92$].  Note that this fitting method is highly uncertain 
for systems with a Type 1 AGN (identified by broad emission 
lines arising from the broad-line region) for which the central engine may contribute 
enough photons to affect the broad-band photometry and alter the SED fitting results.
However, these broad-line AGNs are easily identified and are not the object of this study, 
which targets narrow-line objects (Type 2 AGNs, LINERs, and star-forming galaxies).

We note that, for SDSS galaxies, the method used by Salim and collaborators to derive stellar masses
was tested against the results from \citet{kau03a}. \citet{sal05} found a very good 
agreement between the two distinct methods and calculated the scatter of the difference 
to be 0.11~dex (without 3$\sigma$ outliers, see their Fig.~1), which is smaller than the 
typical uncertainties for our galaxy sample.
Thus, we do not anticipate strong systematic differences to be associated with the
methods used to derive stellar masses for the low-redshift and intermediate-redshift samples
that we use.

In general, we find that the values of stellar mass correlate well with 
the absolute rest-frame $K$-band magnitudes ($M_{K}$).  The latter are obtained 
by applying a $k$-correction to the observed IRAC 3.6$\mu$m photometry.
We calibrate the relation between M$_{\star}$ and $M_{K}$ (see Appendix~\ref{app:mass}) 
for galaxies with both of these estimates in order to estimate a stellar 
mass for galaxies lacking SED fitting calculations.  

Starting from 2,561 galaxies with a valid stellar mass from SED fitting ($\chi^2<$7) 
and satisfying our emission-line selection, we augment our sample with 251 stellar 
masses estimated from $M_K$.  The added galaxies had either missing photometry or an
unacceptable SED fit ($\chi^2>$7). We obtain a total sample size of 2,812 galaxies 
at $0.3<z<1$ with stellar mass estimates and valid \oiiilam\ and \hb\ emission line fluxes.

\subsection{X-ray Luminosity and Classification}\label{sec:Xclass}

We convert X-ray fluxes in the hard band (2 $-$ 8~keV; rest-frame 2.6 $-$
16.0~keV at $z=0.3-1$) to rest-frame 2 $-$ 10~keV
luminosities, assuming a power-law spectrum with photon index as
calculated in \citet{ale03} to perform the $k$-correction. The fluxes are 
corrected for Galactic extinction but not for absorption intrinsic to each galaxy.

Our X-ray classification is based on two criteria: {\it (i)} 
$L_{2-10{\rm keV}} > 10^{42}$~erg~s$^{-1}$; {\it (ii)} hardness 
ratio\footnote{Hardness ratio $\equiv$ (H-S)/(H+S), where H and S are the 
number of X-ray counts in the hard (2-8~keV) and soft (0.5-2~keV) bands.} 
$HR >$ -0.1 (which corresponds to photon index $\Gamma<1$).  X-ray 
sources are classified as AGN if they satisfy {\it at least one} of these 
criteria. Otherwise, they are classified as X-ray starbursts.  
We cannot rule out that some AGNs may fail both the luminosity and hardness
criteria due to X-ray absorption or intrinsically weak emission.  Those objects 
are especially interesting in the framework of this study given the difficulty 
in identifying them using solely X-ray observations.  We keep this possibility 
in mind as we will look for AGN signatures at other wavelengths besides X-rays.  

Some galaxies lack a detection in the hard band ($2-8$~keV) but are detected 
in the full band ($0.5-8$~keV).  In these cases, we $k$-correct the full
band fluxes to obtain rest-frame $2-10$~keV luminosities assuming an
index $\Gamma =$1.9.  These galaxies will be marked with different
plotting symbols when using their inferred $L_{2-10~{\rm keV}}$ since those
values may be limits.

Formally, our classification scheme differs from that described in
\citet{bau04}.  Those authors used a different X-ray luminosity
threshold ($L_{0.5-8~keV}>3\times10^{42}$~erg~s$^{-1}$) and also included criteria 
based on inferred Hydrogen column density and the presence of broad 
($>$ 1000~km s$^{-1}$) or high-ionization emission lines in the optical spectrum.  
Here, we aim for a classification based only on X-rays, independent from
optical spectroscopy.  Nevertheless, we note that the resulting
classification is very similar for galaxies that overlap between the
sample from \citet{bau04} and that presented here.

We consider X-ray detections for most of our analysis but we also
calculated X-ray upper limits for GOODS-N galaxies that were selected
based on their \oiiilam\ luminosity. The X-ray upper limits were
calculated following \S3.4.1 of \citet{ale03} and assuming $\Gamma =$1.9.

\section{Diagnostics at Redshift $0.3-1$}\label{sec:hiz}

Now that we have calibrated the MEx diagram at low-redshift with SDSS galaxies,
we apply it to a sample of intermediate redshift galaxies.  Our sample, described in 
\S\ref{sec:sample}, contains 2,812 galaxies at $0.3 < z < 1$ from the GOODS-N and EGS
fields with valid \hb\ and \oiiilam\ measurements.  

\subsection{Comparison with X-ray Classification}\label{sec:MEx_X}
 
\begin{figure*}
\epsscale{1.} \plotone{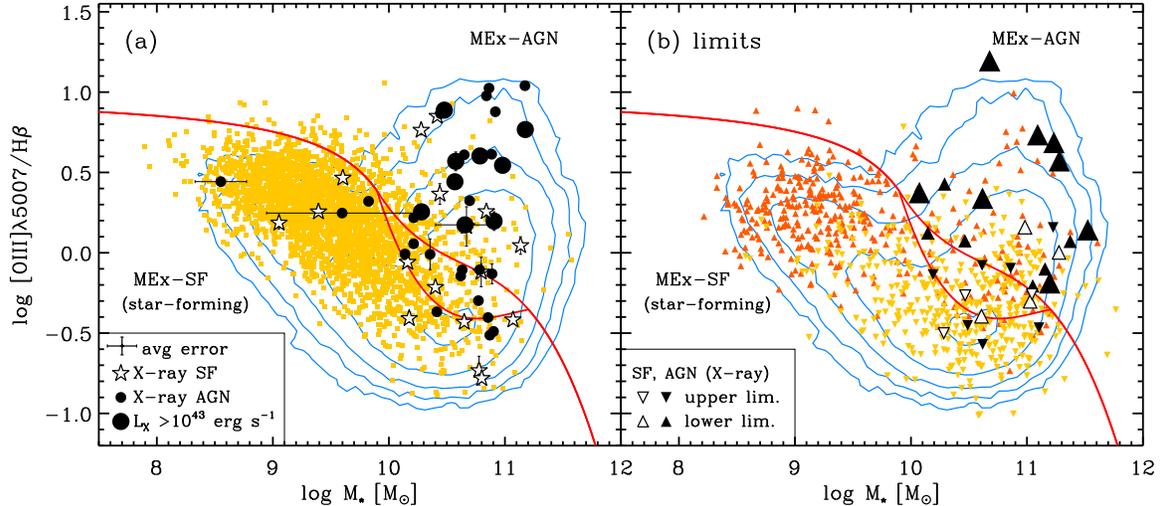}
\caption{Application of the MEx diagnostic to identify the presence of active nuclear 
   activity within galaxies at $0.3<z<1$.  Contours show the SDSS low-$z$ sample 
   (evenly spaced on a logarithmic scale). (a) Our intermediate redshift sample is 
   superimposed (filled squares) and, when available, the X-ray classification 
   is marked with larger symbols [star symbols for X-ray starbursts; small (large) filled 
   circles for X-ray AGNs with $L_X < 10^{43}$~erg~s$^{-1}$ ($L_X > 10^{43}$~erg~s$^{-1}$)].
   We also show galaxies with only one emission line detection and one upper limit (b).
   The resulting \oiii/\hb\ ratios are upper limits when only \hb\ is detected (yellow 
   downward triangles) or lower limits when only \oiiilam\ is detected (orange upward 
   triangles).  The X-ray classification is marked with open triangles for X-ray starburst, 
   and solid black triangles for X-ray AGNs [small (large) for X-ray AGNs with 
   $L_X < 10^{43}$~erg~s$^{-1}$ ($L_X > 10^{43}$~erg~s$^{-1}$)].
   The empirical lines on the MEx diagram are described in the text (see \S\S\ref{sec:bpt},
   \ref{sec:prob}).  The error bar shown in the legend represents the typical uncertainty 
   although we plot individual error bars for objects whose uncertainty are significantly 
   larger.  This diagram is applicable out to $z \sim 1$ and yields spectral classes that 
   are very consistent with the independent X-ray classification shown here.
   (A color version of this figure is available in the online journal.)
   }\label{fig:diagX}
\end{figure*}

The validity of the MEx diagram was demonstrated by showing a good correspondence 
with the BPT classification in \S\ref{sec:bpt}.  Here we show another line of 
support based on the comparison of the MEx classification with a completely 
independent scheme based on X-ray observations.  As can be seen in 
Figure~\ref{fig:diagX}(a), 85\% (34/40) of the X-ray AGNs with valid emission line 
measurements (S/N $>$3) are classified as MEx-AGN (26/40) or MEx-intermediate (8/40) 
on the MEx diagram.  Thus the MEx diagram has a high success rate for recovering X-ray 
identified AGNs.

\begin{figure}
\epsscale{1.} \plotone{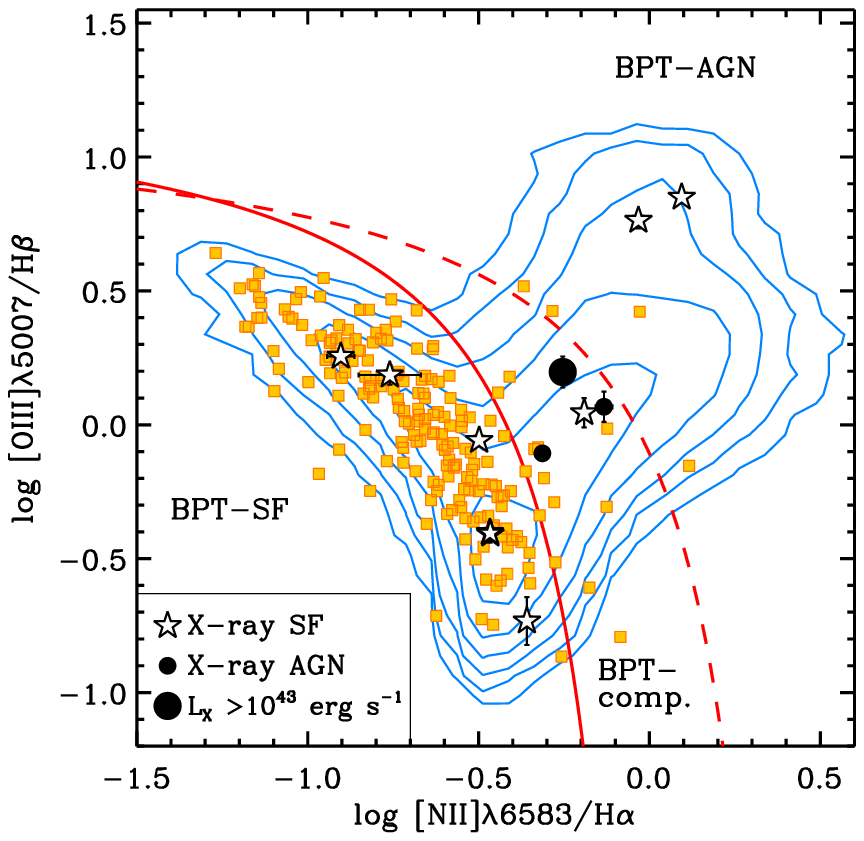}
\caption{
   The BPT diagram, where BPT-AGNs are found above the dashed line \citep[adapted from][]{kew01}, star-forming (BPT-SF)
   galaxies below the solid line \citep[adapted from][]{kau03a}, and composites (BPT-comp.) between the two lines. 
   Contours show the SDSS low-$z$ sample (evenly spaced on a logarithmic scale).
   The $z \sim 0.35$ galaxies for which we have all four diagnostic lines from TKRS 
   ($0.2 < z < 0.5$) and DEEP2 ($0.3 < z < 0.4$) are mostly distributed along the 
   star-forming sequence, with a few galaxies extending along the AGN plume (filled squares). 
   We identify galaxies with a {\it Chandra} detection as a function of their X-ray 
   classification [star symbols for X-ray starbursts; small (large) filled 
   circles for X-ray AGNs with $L_X < 10^{43}$~erg~s$^{-1}$ ($L_X > 10^{43}$~erg~s$^{-1}$)].  
   While all the X-ray identified AGNs lie above the Kauffmann line (solid line), 
   there are also two galaxies nominally classified as X-ray starbursts that lie in 
   the region where AGNs are prominent.  These systems also lie in the MEx-AGN region 
   [Figure~\ref{fig:diagX}(a)] and they may be X-ray weak or X-ray absorbed AGNs.  
   In either case, the optical diagnostic is a useful complement to X-ray observations alone.
   (A color version of this figure is available in the online journal.)
   }\label{fig:bptX}
\end{figure}

\begin{figure*}
\epsscale{1.05} \plotone{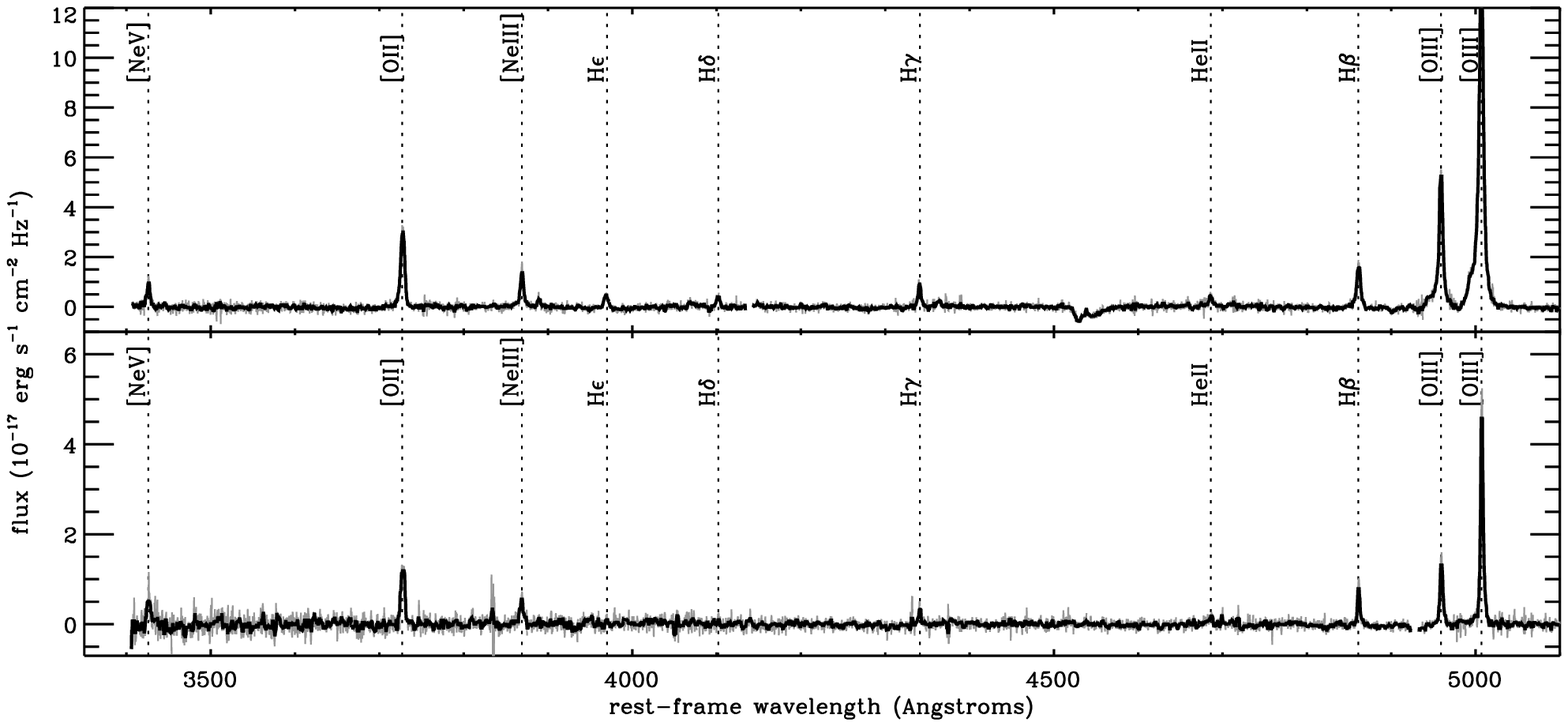}
\caption{
   Two example TKRS spectra with the most obvious \nevlam\ detections.  
   While the top spectrum is for a galaxy that was also identified as AGN 
   in the X-rays (J123608.13+621036.2, $z$=0.679), the bottom one is for a galaxy 
   that was classified as X-ray starburst (J123645.40+621901.3, $z$=0.455).  The 
   spectra are continuum-subtracted (grey line) and smoothed with a running
   median (thicker black line).  Typical emission lines are marked with vertical 
   dotted lines and labeled.
   }
\label{fig:nev}
\end{figure*}

As for the X-ray starbursts, 50\% (8/16) are classified as SF on our new diagram, while 
19\% (3/16) are in the intermediate region and the remaining 31\% (5/16) reside in the AGN
region.  However, low-luminosity AGNs and some heavily absorbed AGNs may appear as faint 
as X-ray starbursts so this class of objects likely includes these different systems in 
addition to genuine starbursting galaxies.  
Indeed, the two X-ray starbursts that are in the AGN region of the MEx diagram but that lie 
at sufficiently low redshift to be placed on the BPT diagram ($z<0.5$) stand out in the BPT-AGN region
(Figure~\ref{fig:bptX}), further confirming the presence of actively accreting black holes in these galaxies.
The optical diagnostics are thus especially useful when the X-ray signal alone is ambiguous 
(e.g., too faint to securely identify AGNs). 
Additional support for the presence of AGN in the X-ray starburst class 
is provided for at least one galaxy for which we have a clear detection of the 
\nevlam\ emission line (Figure~\ref{fig:nev}).  This transition is an 
unambiguous tracer of AGN activity because of it's high ionization 
potential (97.1\,eV).

Next, we use the MEx classification probability method 
described in \S\ref{sec:prob}.  We calculate $P(AGN)$ by adding the 
probabilities of any AGN category (composite, LINER and Seyfert 2).  
We will show in \S\ref{sec:Xstack} that $P(AGN)>30\%$ is a useful 
threshold to distinguish AGNs from purely star-forming galaxies.  
When X-ray AGNs, X-ray starbursts, and non-X-ray detections are 
considered separately,
we find that 29 among 35 X-AGNs (83\%) galaxies have P(AGN)$>$30\%, 
with an average AGN classification probability of 77\%.  For the X-ray starbursts, 
which all have a low hard X-ray luminosity  ($L_{2-10{\rm keV}} < 10^{42}$~erg~s$^{-1}$), 
we find that 11 among 17 X-SBs (65\%) have P(AGN)$>$30\% (average P(AGN)=52\%). 
This suggests that some of them host an X-ray absorbed or X-ray weak AGN.  
Lastly, we note that some of the galaxies lacking an X-ray detection lie well into 
the AGN region with P(AGN) up to 100\%. These X-ray faint AGN candidates are studied in more 
detail in \S\ref{sec:Xstack}.  Like the X-ray starburst class, they may in fact 
include weak or heavily absorbed AGNs.

Not all of the X-ray detected sources have valid emission line measurements.  X-ray galaxies
for which only one of \hb\ or \oiiilam\ is measured and the other line has an upper limit
are shown in Figure~\ref{fig:diagX}(b).  In these cases, the \oiii/\hb\ line ratio is either 
an upper or lower limit.  We find that most X-ray classified AGNs appear consistent with their
optical classification given their limits on the MEx diagram.  X-ray galaxies for which no lines
could be measured due to insufficient data quality or intrinsic weakness of the emission lines
are considered in \S\ref{sec:Xnoline}.

We note that the locus of our intermediate-redshift star-forming galaxies may be offset 
slightly from the SDSS contours.  A more careful analysis of the selection effects would be 
required to determine whether such a shift is real or simply results from selection 
biases.  We discuss possible evolutionary effects in \S\ref{sec:evol}.

Overall, the MEx diagram is a new tool that can be used on galaxies with 
\oiiilam\ and \hb\ line fluxes as well as stellar mass.  It is thus possible to 
apply a robust AGN diagnostic to optical spectroscopic samples out to $z \sim 1$, 
provided the galaxies also have photometry allowing a stellar mass estimate.  
An independent X-ray classification scheme supports the validity of our diagram. 
This means that surveys with optical spectroscopy but lacking X-ray coverage can
still benefit from a reliable AGN diagnostic.

In order to complete the comparison between the optical MEx diagram AGN classification, 
and the X-ray based classification, we perform two experiments.  First, 
we use X-ray stacking to search for AGN signatures in galaxies that are 
undetected in the X-ray observations (\S\ref{sec:Xstack}).
Second, we calculate the demographics of all X-ray AGNs regardless of the presence of optical 
emission lines (\S\ref{sec:Xnoline}).  This exercise highlights the complementarity of the 
optical and X-ray selection techniques.

\subsection{X-ray Stacking}\label{sec:Xstack}

Stacking the X-ray observations allows us to probe X-ray emission to fainter flux 
levels and, in the case of significant detections in more than one energy band, to 
estimate the X-ray spectral slope. The X-ray spectral slope (or photon index $\Gamma$) 
can be used to distinguish between different source types.  While unabsorbed AGNs 
and low-mass X-ray binaries (LMXBs) exhibit a large index ($\Gamma>1.3$ and 
$>1.7$, respectively), high-mass X-ray binaries (HMXBs) can yield flatter slopes with 
indices down to ($\Gamma>0.5$) and absorbed AGNs have still flatter slopes with 
($-1<\Gamma<1$).  These ranges are illustrated on Figure~2 of \citet[][see references therein]{ale05}.  

In the case of X-ray absorbed AGNs, the flattening is caused by the photoelectric 
absorption of soft X-ray photons by heavy atoms in neutral gas along the line of sight.  
Because softer photons are preferentially absorbed relative to harder ones, Compton-thick 
AGNs (with $N_H>10^{24}~{\rm cm}^{-2}$) should exhibit a flat slope with $\Gamma<1$. 
However there are at least two situations where Compton-thick AGNs instead have a steep 
X-ray slope at energies $<10$~keV.  First, the shape of the X-ray spectrum varies 
greatly when the X-ray emission is dominated by reflection rather than direct transmission.  
In this case, the slope of the spectrum does not truly reflect the hardness of the 
radiation and one must use other indicators to estimate X-ray absorption (e.g., 
the equivalent width of the Fe~$K\alpha$ line).  
Second, the photon index at energies $0.5-10$~keV can be significantly altered when 
soft emission from starburst activity is coincident with the 
harder emission from an absorbed AGN.  We keep these possibilities in mind 
when we interpret the results of X-ray stacks.

In this paper, we perform two different X-ray stacking analyses.  The first 
analysis is a proof-of-concept of the MEx diagram.  In this case, the 
goal is to stack the X-ray signal for subsamples 
defined from the MEx classification.  A more detailed description and 
the results are provided in this section.  For the second analysis 
(described below in \S\ref{sec:Xstack_Cthik}), we apply additional criteria 
to target specifically X-ray absorbed AGN candidates.

In this section, we are interested in galaxies that are not detected
individually in the deep 2~Msec {\it Chandra} observations in GOODS-N
but that have the required emission line measurements for the MEx diagram.
Employing a stellar mass cut-off at 10$^{10}~M_{\sun}$ and requiring 
\hb\ and \oiiilam\ emission line detections (S/N$>$3), there are 69 such 
galaxies at 0.3$<z<$1.

For the X-ray stacking we adopt a similar approach to \citet{wor05}, 
who stacked sources and calculated the significance of the stacked result 
using 10,000 Monte Carlo trials. In all cases, we limit the X-ray stacking
to sources within 6~arcmin of the {\it Chandra} aim point to maximize
sensitivity and we stack the sources in the soft and hard
bands. Sources that lie within a factor of 2 of the 90\% encircled
energy radius of another X-ray detected source are excluded. We only
consider the stacked signal as a significant detection when the
number of detected counts in a given band exceed the background
count rate determined from the Monte Carlo trials by $\ge$~3~$\sigma$.

Following the criteria outlined above there are 47 (out of 69) X-ray 
undetected galaxies in GOODS-N for which we can stack the {\it Chandra} 
data. We select subsamples of the X-ray undetected galaxies to stack based on
the probabilities $P(AGN)$ \& $P(SF)$. Here we use $P(AGN)$ as the total
probability for the composite, LINER and Seyfert subclasses. 
We divide the sample at $P(AGN)>50$\% and also at $P(AGN)>30$\% in order to 
include the MEx-intermediate region.

In the former case, we find significant detections in {\it Chandra}'s soft 
band and marginal detections in the hard band for these $P(AGN)>50$\% and 
$P(SF)>50$\% subsamples of 25 and 22 galaxies, respectively.  The X-ray spectral 
slopes are respectively flat ($\Gamma\sim0.8$) and slightly steeper 
($\Gamma\sim1.2$; Table~\ref{tab:X}).  A flat X-ray spectral slope of $\Gamma\sim$~0.8 
unambiguously indicates the presence of absorbed AGN activity 
\citep[see Fig.~2 of][]{ale05,mai98} in the P(AGN)$>$50\% 
subsample. The stacked signal of the complementary subsample may correspond 
to a mixture of star-forming galaxies and some absorbed AGNs.  However, the 
difference in $\Gamma$ between the $P(AGN)>50$\% and $P(SF)>50\%$ subsamples is 
not statistically significant so there is a fair likelihood that some absorbed 
AGNs are present.  This result suggests 
that P(AGN)$>$50\% may not be sufficient to recover all of the AGNs.

Imposing a lower cutoff at $P(AGN)>30$\% yields 34 objects to stack. Stacking
these sources gives significant detections in the soft and hard bands
corresponding to a flat X-ray spectral slope of $\Gamma\sim$~0.8,
again unambiguously indicating the presence of absorbed AGN activity. 
We then stacked the remaining 13 systems, with $P(SF)>70$\%.  In this case, 
we only find a significant detection in the soft band.  The steep X-ray 
spectral slope ($\Gamma\sim1.7$) is consistent with the X-ray emission of 
these galaxies being dominated by star formation processes.  
These analyses are summarized in Table~\ref{tab:X} and provide good 
first-order confirmation that the combination of the MEx diagram and our 
probabilistic approach provides a comparatively clean selection of star-forming 
galaxies and AGNs.


Our results suggest that using $P(AGN)>30$\% as a threshold for nuclear
activity leads to a cleaner separation between AGNs and star-forming galaxies
than using $P(AGN)>50$\%.  We examine the objects with intermediate AGN likelihood 
separately by stacking the 12 galaxies with $30\%<P(AGN)<50\%$. 
As expected, galaxies in this subset are likely composite systems.
We calculate $P(comp)$, $P(LINER)$, $P(Sy2)$ separately and, on average,   
the composite class is 5.5 times more likely than the LINER and Seyfert~2 classes 
taken together.  We find a flat spectral slope of $\Gamma\sim$~0.6, statistically 
indistinguishable from the P(AGN)$>$50\% subsample.  This result indicates that 
some galaxies with $30\%<P(AGN)<50\%$ host an X-ray absorbed AGNs, therefore 
are an important population to search for Compton-thick AGNs and to take into 
account for a complete census of AGNs.

We have demonstrated that the MEx diagnostic diagram works well at intermediate 
redshift ($0.3<z<1$) given that the majority of the X-ray AGNs lie in the 
MEx-intermediate and MEx-AGN regions rather than the MEx-SF region.  The X-ray 
stacking analyses presented in this section consolidate this result.  
Furthermore, there are unidentified AGNs within the X-SB class (nominally X-ray 
starburst although some sources with faint luminosities are more difficult to 
classify unambiguously) and within an X-ray undetected population.  These can be 
identified using the MEx diagram.

\subsection{Demographics of X-Ray Selected AGNs}\label{sec:Xnoline}

So far, we have compared the X-ray and MEx classification schemes for galaxies with
valid emission line measurements (S/N$>$3 for \hb\ and/or \oiiilam).  Here, we also 
consider X-ray detected sources that have sufficient spectral coverage for both emission 
lines but nonetheless lack detections.  We use the X-ray AGN classification 
described in \S\ref{sec:Xclass} to keep only secure X-ray AGNs (i.e., with 
$L_{2-10keV}>10^{42}$~erg~s$^{-1}$ or hardness ratio $HR>$-0.1).

Combining GOODS-N and EGS subsamples, there are 101 X-ray detected AGNs with \hb\ and 
\oiiilam\ within the spectroscopic wavelength coverage. We require S/N$>$3 for either
\hb\ or \oiiilam\ in order to perform the MEx classification.  Based on this criterion, 
33\% of the X-ray selected AGNs lack optical emission line signatures.  
Performing the MEx classification for the remaining 68 objects with emission line 
measurements shows that 60 are in the MEx-AGN or MEx-intermediate region, and 
the remaining 8 lie in the MEx-SF region.  To summarize the demographics, 
59\% are in the MEx-AGN or MEx-intermediate regions (48\% and 11\%, respectively), 
8\% are in the star-forming region, and 33\% lack a classification
because their emission lines are either too weak or corrupted.

As a comparison, Y11 used an X-ray selected sample in EGS that is a 
superset of the EGS sample studied here.  They combined DEEP2 spectra and 
MMT spectra that were obtained specifically to target X-ray sources 
that were not part of the DEEP2 slit masks \citep{coi09}.  
For their X-ray selected sample, they found that 51\% met their AGN class 
based on the color-excitation diagram, 22\% were in the star-forming region of their 
diagram and 25\% lacked emission lines.  Yan and collaborators refer to the latter as 
X-ray bright, optically-normal galaxies (XBONGs), which they study in more detail 
\citep[also see][]{rig06,tru09}.

A distinction between the present work and the results presented in Y11 is their 
lower signal-to-noise cutoff for emission line measurements (S/N$>$2).  We prefer 
to use a more conservative selection by imposing S/N$>$3 based on the visual inspection 
of the spectra and the large residuals often present near the lines with S/N$<$3.  
Allowing for a lower S/N is likely responsible for the apparent smaller fraction 
of galaxies lacking emission line measurements in Y11 although the difference between 
the two analyses is not statistically significant ($\sim1\sigma$).  We stress that using the  
intermediate region -- where composite and star-forming galaxies overlap -- 
helps to recover a larger fraction of the X-ray detected AGNs, 
indicating the importance of composite systems in achieving a global AGN census.

In summary, we find that X-ray selected AGNs cover the full range of optical 
spectroscopic classes (AGN, SF, no or weak emission lines). Our results are similar 
to previous work.  Also, we have shown that the MEx and X-ray AGN selection methods remain 
complementary.  
Some  AGNs lack X-ray detection (or are X-ray faint) and we found evidence for absorption among 
that population.  Given the importance of X-ray absorbed AGNs, and the possibility to 
start identifying this missing population, we will now focus on this topic.

\section{Compton-Thick AGN Candidates}\label{sec:AGNpower}

While soft (0.5$-$2~keV) X-ray emission may be produced by star-formation 
activity, a large amount of hard (e.g., 2$-$10~keV) radiation is considered 
to be an obvious signature of AGN activity.  However, large column densities 
($N_{H} > 10^{24}$~cm$^{-2}$) can cause severe absorption even for hard X-rays, 
making it challenging to identify Compton-thick AGNs even with sensitive 
{\it Chandra} observations such as those used in this work.

In the event that photons with energies between 2$-$10~keV are absorbed by 
intervening material, it is possible to estimate the intrinsic luminosity 
of an AGN by looking at emission originating from larger scales than the 
source of absorption \citep[e.g.,][]{mul94,bas99}.  Because \oiiilam\ emission 
arises in the narrow-line region, which extends several hundred parsecs beyond 
the active nucleus, it is not subject to the same small-scale nuclear absorption 
as hard (2-10~keV) radiation.  This type of compact absorption is expected 
for geometries involving an obscuring torus as suggested in AGN unification 
models \citep{ant93}. Thus, the ratio of X-ray (2-10~keV) to intrinsic (dust 
corrected) \oiii\ luminosities can serve as a Compton-thickness parameter 
$T$ [$\equiv$ L(2-10~keV)/L(\oiiilam)] \citep{bas99}.  This quantity has 
two main caveats.

The first caveat is the uncertain obscuration of \oiiilam\ by dust in the 
host galaxy \citep[external to the narrow line regions,][]{mai95}.  For 
example, \citet{lama09} found a median correction of 1.0~mag (0.5 to 2.3 mag) 
at \oiiilam\ based on the Balmer decrement using a SDSS-selected 
sample of 17 Seyfert~2 galaxies that are \oiii-luminous.  Given the 
uncertain dust obscuration corrections for \oiii, we conservatively choose 
to use the observed luminosities in our analysis while keeping in mind 
that a typical correction of 1.0~mag would shift the values of the 
Compton-thickness parameter down by $\sim$0.4~dex.

The second caveat is the possible contribution of stellar photoionization 
to the \oiii\ luminosity.  \citet{kau03c} calculated the fractional 
contributions to emission line luminosities in galaxies hosting both star 
formation and AGN activity.  For metal-rich galaxies with AGN, they found 
that only around 7\% of \oiiilam\ luminosity is due to star formation
(the remaining 93\% originates from AGN-excited gas).  This is in contrast 
with other lines such as \hb\ and \oiilam\ for which a larger contribution 
can be expected to come from \hii\ regions rather than the narrow line 
region photoionised by the AGN (up to 45\%$-$70\%).

However, the stellar contribution to $L_{\oiii}$ can be much more significant 
in low-metallicity galaxies since oxygen becomes one of the main coolants and
changes in line blanketing produce a harder ionization field \citep{kew06}.  
To alleviate this source of contamination to the \oiii\ luminosity, we restrict the  
Compton-thickness analysis to high stellar mass (and presumably high-metallicity) 
galaxies, with $M_{\star}>10^{10.2}~M_{\sun}$.  

\subsection{X-Ray Absorption Versus Observed AGN Power}\label{sec:TLx}

\begin{figure}
\epsscale{1.} \plotone{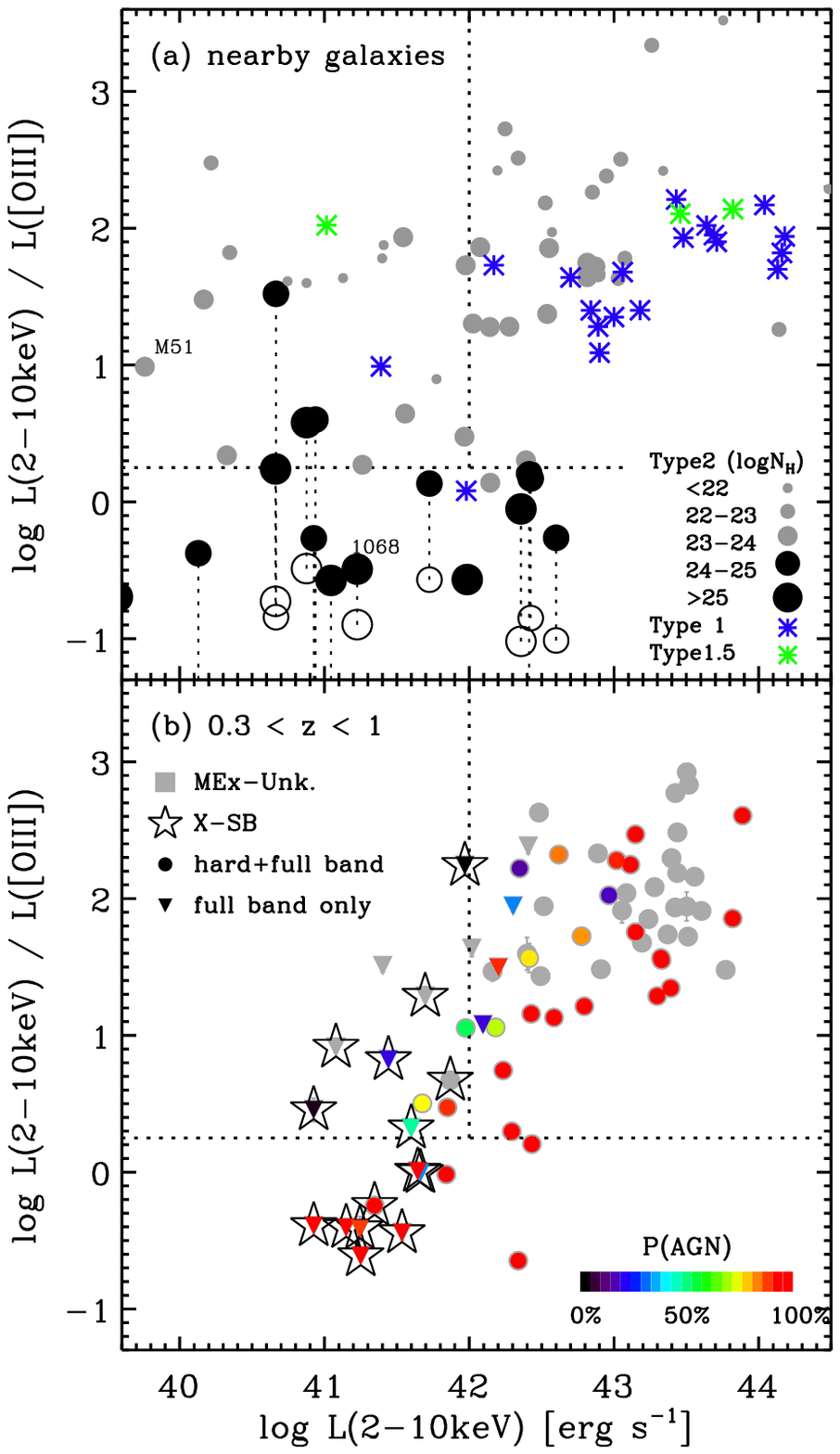}
\caption{
   Compton-thickness parameter $T \equiv L_{2-10keV}/L_{\oiii}$ as a function of hard 
   X-ray luminosity ($L_{2-10keV}$) for (a) nearby galaxies and (b) 
   intermediate-redshift galaxies. In panel (a), the \oiii-selected type~1 AGN 
   sample from \citet{hec05} is shown with blue asterisks.  Type~1.5 and type~2 AGNs 
   (green asterisks and filled circles, respectively) are taken from \citet{bas99}.  
   The symbols are keyed to the column density $N_H$ in the case of type~2
   AGNs (see legend).  Dust-corrected \oiii\ values are shown with open circles for galaxies 
   with $N_H > 10^{24}\,{\rm cm}^{-2}$.  The intermediate-redshift sample (b) is selected to have 
   log($M_{\star}$[$M_{\sun}$])$>$10.2, and is color-coded according to the probability to host 
   an AGN from the MEx diagram (see color bar).  The very likely AGNs (orange and red) 
   tend to follow the bottom envelope of the points, and reveal some Compton-thick AGN 
   candidates in galaxies that were otherwise classified as X-ray starbursts (star symbols; 
   see \S\ref{sec:Xclass}).  
   Most X-ray AGNs (points lacking star symbols) are characterized by log($L_X/L_{\oiii}$) $>$ 1.
   Galaxies that were not identified on the MEx diagram (grey points) fail the emission-line 
   quality flag for \hb.
   }\label{fig:OiiiXM}
\end{figure}

In order to compare the X-ray absorption to the observed AGN power, we plot the 
Compton-thickness parameter $T$ as a function of the observed (i.e., not corrected 
for absorption) hard X-ray luminosity (Figure~\ref{fig:OiiiXM}).  

\citet{hec05} found that the average values of log($T$) are 1.59$\pm$0.48~dex and 
0.57$\pm$1.06~dex for Type 1 and Type 2 AGNs, respectively.  Type 1s are less 
absorbed and show a tighter correlation while Type 2s include a much broader range of 
X-ray absorption (going down to smaller values of $T$).  We reproduce this result
in Figure~\ref{fig:OiiiXM}(a) where we combine samples of Seyferts from \citet{hec05}
and \citet{bas99}.  We also see a clear trend between the location 
of nearby Seyfert 2's on Figure~\ref{fig:OiiiXM}(a) and their inferred column density
$N_H$ from \citet{bas99}.  As expected, the Compton-thick galaxies, with $N_H > 10^{24}~{\rm cm}^{-2}$,
tend to lie at low values of $T$ [log($T$)$\lesssim$0.25 for observed X/\oiii\ ratios, 
and log($T$)$<$-0.3 after correcting $L_{\oiii}$ for dust obscuration].

Turning our attention to the sample of $0.3 < z < 1$ galaxies, we find a similar trend 
in X-ray absorption versus hard X-ray luminosity as found for the nearby galaxies.  There is an 
apparent transition at $L_{X} \approx 10^{42}$~erg s$^{-1}$ (vertical dotted line).  X-ray AGNs 
above that luminosity threshold span a restricted range in their X-to-\oiii\ luminosity 
ratio ($1<\log(T)<3$) compared to fainter hard X-ray sources.
Galaxies that likely host an AGN based on the MEx diagram lie along the lower envelope
of the distribution, consistent with their large \oiii/\hb\ ratios being driven by 
luminous \oiii\ emission.  Galaxies that are likely star forming according to the MEx diagram 
have lower X-ray luminosities on average, and occupy the mid-range of the 
Compton-thickness parameter values.

While some of the X-ray starbursts (open star symbols) were also classed as SF from 
the MEx diagram, there are a few objects with a conflicting spectral class.  These X-ray-SB 
but MEx-AGN galaxies (open stars around red circles) are clustered around $\log(T)=0$, 
the nominal value for Compton-thick AGNs.  Given the predicted offset due 
to using uncorrected \oiiilam\ luminosities, and the location of known nearby 
Compton-thick AGNs on Fig.~\ref{fig:OiiiXM}(a), we will adopt $\log(T)<0.25$ as a criterion 
for identifying Compton-thick AGN candidates in the next sections. 
In cases with low values of $\log(T)$ the \oiiilam\ line luminosity may be a better 
indicator of the bolometric AGN luminosity.  We investigate this next.

\subsection{X-Ray Absorption Versus Intrinsic AGN Power}\label{sec:TLOiii}

In the previous section, we considered candidate absorbed AGNs with a 
detection in the X-rays, at least in the full band.  In this section, we 
will expand by including cases with upper limits in X-rays.  Limits were 
derived for observations in GOODS-N as described in \S\ref{sec:Xclass}.

\begin{figure}
\epsscale{1.} \plotone{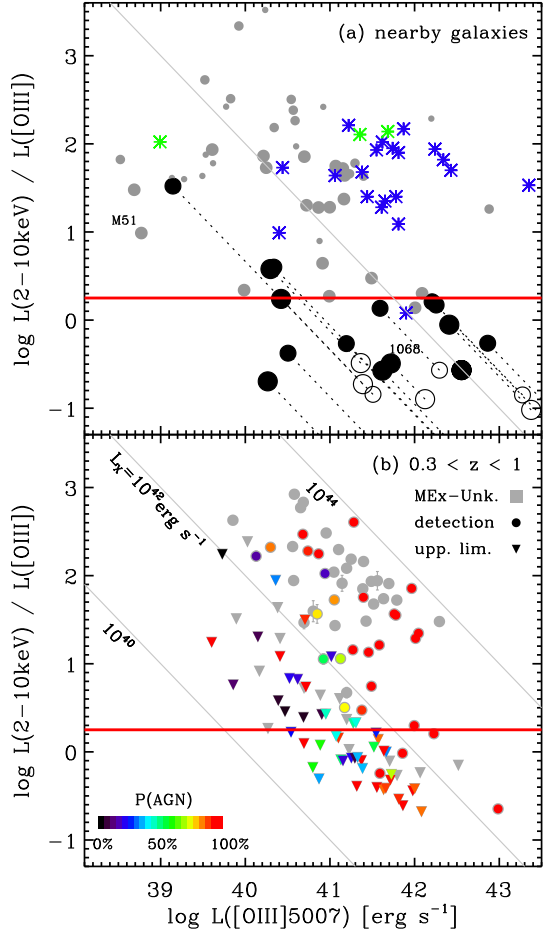}
\caption{
   Compton-thickness parameter $T \equiv L_{2-10keV}/L_{\oiiilam}$ as a function 
   of \oiiilam\ luminosity.  Both panels are shown on the same scale. 
   Nearby galaxies are plotted in panel (a) and the symbols are as described 
   in Fig.~\ref{fig:OiiiXM}.  The intermediate-redshift sample is shown 
   in panel (b).  The diagonal lines correspond to fixed hard X-ray ($2-10$ keV) 
   luminosities of $10^{40}, 10^{42}, 10^{44}$~erg s$^{-1}$ as labeled.  
   We note that $L_{2-10keV}=10^{42}$ ~erg s$^{-1}$ is the nominal division that we adopt 
   to separate X-ray AGNs from the X-ray starbursts although there are some 
   X-AGNs below that luminosity threshold (recognized by their hardness ratio 
   $HR>-0.1$). In addition to the X-ray classification, the MEx AGN probabilities are 
   indicated with the color coding.  P(AGN) includes all AGN classes, such 
   that P(AGN)$+$P(SF)$=$100\%.  Galaxies that were not identified on the MEx 
   diagram (grey points) fail the emission-line quality flag for \hb.  
   This figure shows that star-forming/starburst galaxies occupy a 
   contiguous region with the absorbed AGN candidates, stressing the 
   importance of the MEx diagram in selecting likely AGNs.  Based on the 
   observed position of the known Compton-thick galaxies in panel (a), we 
   adopt log($L_X/L_{\oiii}$)$=$0.25 as a criterion for Compton-thick 
   candidates (horizontal line).      
}\label{fig:TLOiii}
\end{figure}

We again utilize the Compton-thickness parameter in Figure~\ref{fig:TLOiii}.  
Here, we use $L_{\oiii}$ on the horizontal axis to probe the intrinsic 
AGN power.  Nearby \oiii-selected Seyfert 1's from \citet{hec05} occupy 
the bright (elevated $L_{\oiii}$) and unabsorbed (log($T$)$>$1) region 
of Figure~\ref{fig:TLOiii}(a).  Seyfert 2's span a broad range of X-ray 
absorption and intrinsic AGN power.

In our intermediate-redshift sample (panel b), we note that star-forming 
galaxies seem to occupy a region contiguous with the Compton-thick candidates, 
which have log($L_{2-10 keV}$)/log($L_{\oiii}$)$<$0.25.  To discard these 
systems and keep absorbed AGN candidates, we select galaxies below the 
$L_X/L_{\oiii}$ threshold (red line) that also have an significant probability 
of hosting an AGN according to the MEx diagram.  We adopt P(AGN)$>$30\% and 
stack the {\it Chandra} observations to search for hidden AGN signal in 
\S\ref{sec:Xstack_Cthik}.

Figure~\ref{fig:TLOiii} shows that the Compton-thick AGN candidates, with 
$\log(T)<0.25$, span a similar range in \oiiilam\ luminosity as the 
unabsorbed AGNs, with $\log(T)>1$. 
At high \oiiilam\ luminosities (e.g., $L_{\oiiilam}>10^{41}$erg~s$^{-1}$) 
the distribution of Compton-thickness values appears to be bimodal, 
with the AGNs being either largely unabsorbed ($\log(T)>1$) or heavily 
absorbed ($\log(T)<0.25$).  We note that this possible bimodality 
is seen in both the nearby and the intermediate-redshift galaxy 
samples. Our sample is mostly complete at the bright \oiii\ 
end\footnote{We detect galaxies with $L_{\oiiilam}>10^{41}$erg~s$^{-1}$ 
out to $z=0.95$} 
so we do not believe that this trend can be attributed to selection 
effects.  However, the number statistics are too small to investigate 
it further.

Unsurprisingly given the fairly small volume probed, we mostly detect AGNs 
with Seyfert-like luminosities and below [$\log(L_{\oiii})<42.5$~erg~s$^{-1}$].
However, we note two possible AGNs with intrinsic 
luminosity in the quasar regime [with $\log(L_{\oiii})>42.5$~erg~s$^{-1}$], both 
of which appear to suffer from large X-ray absorption.  These systems may be 
important testbeds for evolutionary scenarios where quasars are born 
in a deeply enshrouded environment (presumably following a major galaxy merger) 
before a blowout phase where the surrounding material is ejected revealing the 
optically and X-ray bright central engine \citep{san96,dim05,hop05}.  We 
leave this analysis for future work.

In this section we have investigated the presence of X-ray absorbed AGNs 
by using both X-ray detected and X-ray undetected \oiii-selected 
objects.  We find that the range of intrinsic luminosities probed by $L_{\oiii}$ 
are comparable for Compton-thick candidates and X-ray unabsorbed systems.  The 
presence of X-ray  absorption is inferred from the X-to-\oiii\ luminosity ratio ($T$) and we 
utilize X-ray stacking analyses in the following section to further justify the presence
of absorption in these systems.

\subsection{X-ray Stacking of Highly-Absorbed AGNs}\label{sec:Xstack_Cthik}

Here we search for X-ray signatures in the X-ray undetected absorbed AGNs
with low X-ray/\oiii\ ratios identified in the previous section
(\S\ref{sec:TLOiii}). There are 33 galaxies that are part of our candidate 
Compton-thick AGN selection\footnote{log($L_{2-10 keV}$)/log($L_{\oiii}$)$<$0.25, 
P(AGN)$>$30\% and log($M_{\star}[M_{\sun}]$)$>$10.2.} (see Fig.~\ref{fig:TLOiii}b). 
Sixteen are X-ray detected and are examined on a case-by-case basis below. Of 
the 17 X-ray undetected objects, 13 lie sufficiently close to the {\it Chandra} 
aim point without lying too close to X-ray bright sources to allow for X-ray 
stacking analyses.  We employ the same X-ray stacking method described in \S\ref{sec:Xstack}.  
However, while in \S\ref{sec:Xstack} we were concerned with stacking the 
X-ray data for galaxies selected solely from the MEx classification scheme, 
galaxies in this section are selected as potential Compton-thick AGNs by 
requiring $\log(T)<0.25$. 

Of the 16 X-ray detected sources, six are bright at X-ray energies 
($F_{0.5-8keV}>$3$\times$10$^{-16}$~erg\,s$^{-1}$\,cm$^{-2}$) and ten are
faint ($F_{0.5-8keV}<$3$\times$10$^{-16}$~erg\,s$^{-1}$\,cm$^{-2}$).
The galaxies with bright X-ray fluxes tend to lie at 
lower redshift and have steep spectral slopes ($\Gamma\approx$~1.3--2.0).  
Two of the six X-ray bright sources have a clear \nevlam\ detection
(Fig.~\ref{fig:nev}), an unambiguous tracer of AGN, and reside in the Seyfert 2 
region of the BPT diagram.  However, only one of them is clearly identified 
as an X-ray AGN ($L_X>10^{42}$\,erg\,s$^{-1}$; J123608.13+621036.2 in Fig.~\ref{fig:nev}). 
This object is strongly absorbed ($\Gamma \sim$0.2) and is a potential 
Compton-thick AGN.  The remaining five X-ray bright sources fail both the 
luminosity and hardness criteria to be clearly identified as X-ray AGNs.

As for the X-ray faint galaxies, one is detected in the hard band and has 
a flat X-ray spectral slope ($\Gamma\approx$~0.7), indicating that it hosts 
a heavily absorbed AGN.  The other nine objects are too faint to provide 
significant constraints on their X-ray spectral slopes. However, from stacking 
the {\it Chandra} data of these nine X-ray faint galaxies we find a comparatively 
steep X-ray spectral slope ($\Gamma\approx$~1.4), consistent with that found for 
the X-ray bright galaxies. 

We stacked the {\it Chandra} data for the 13 X-ray undetected Compton-thick 
AGN candidates, which we treat separately from the individual detections. 
While the individually detected galaxies show steep X-ray indices, for 
the stacked objects we find a very flat photon index of $\Gamma\approx$0.4 
(see Table 2).  The flatter slope found here strongly suggests the presence of 
X-ray absorbed AGN activity in at least a fraction of the objects that were stacked.  
The combined low X-ray photon index and small values of the thickness parameter 
make these systems robust Compton-thick AGN candidates.

\begin{figure}
\epsscale{1.05} \plotone{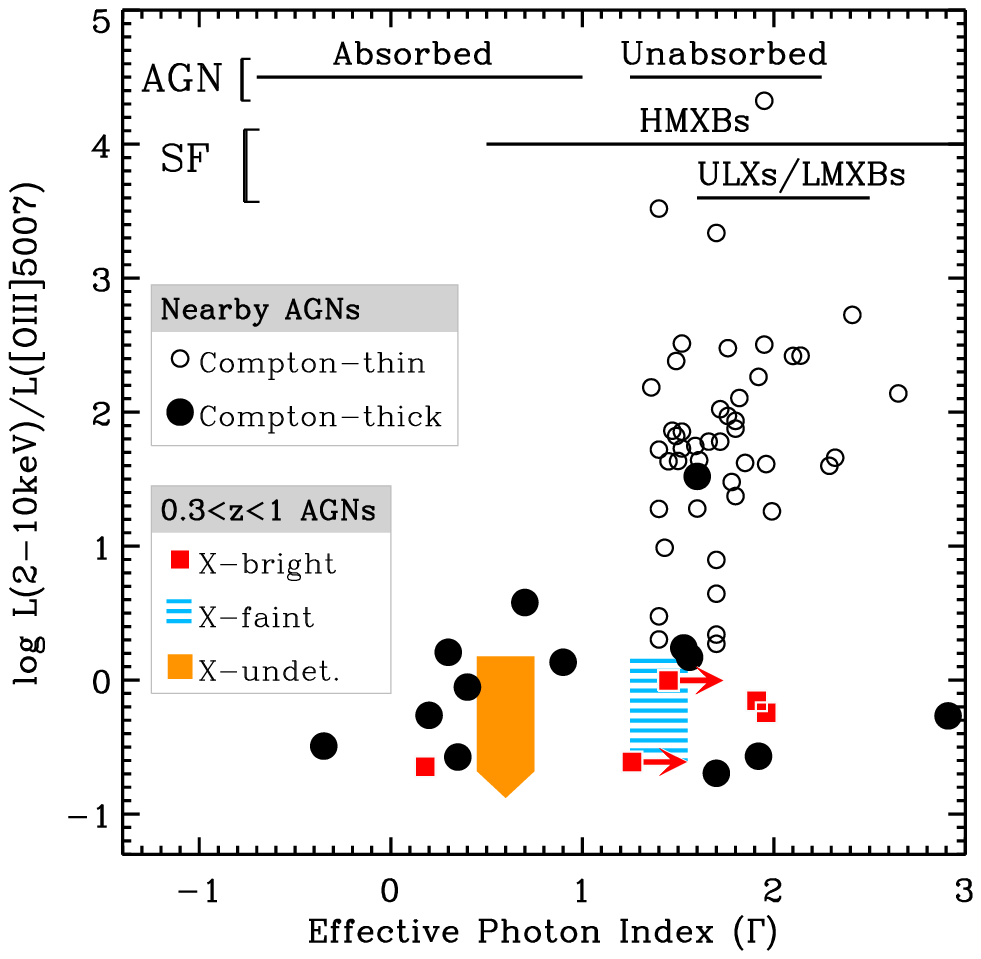}
\caption{Compton-thickness parameter $T \equiv L_{2-10keV}/L_{\oiii}$ as a function 
   of the effective photon index ($\Gamma$) calculated as described in \citet{bas99} 
   and \citet{ale03} for nearby and intermediate-redshift galaxies, respectively. 
   The expected ranges of $\Gamma$ are shown for various sources of X-ray emission 
   at the top of the figure \citep[taken from][and references therein]{ale05}. 
   While Compton-thin galaxies have $\Gamma>1$ as expected (open circles), 
   Compton-thick AGNs can have either a flat or a steep index (filled circles; 
   see text for details).  We compare the locus of these nearby AGNs (open and 
   filled circles) with the $0.3<z<1$ systems that we identified as Compton-thick 
   AGN candidates based on the following criteria: $\log(T)<0.25$, P(AGN)$>$30\% 
   and $\log(M_{\star}[M_{\sun}])>10.2$.  The intermediate-redshift galaxies with 
   bright X-ray fluxes (red squares) span a range of photon indices.  One galaxy 
   has an obviously flat slope ($\Gamma \sim 0.2$) and the remaining four have 
   steeper slopes ($\Gamma > 1$) but are consistent with the range of values 
   spanned by known, nearby Compton-thick AGNs \citep[solid circles, from][]{bas99}.
   We furthermore show the results from X-ray stacking of candidate Compton-thick 
   AGNs that are either X-ray weak (light blue hatched region) or X-ray undetected 
   (orange shaded region).  The height of the shaded regions illustrates the range 
   of values for the individual galaxies used in the stack (upper limits in the 
   case of the non-detections).  The sample of X-ray non-detections clearly includes 
   absorbed AGNs.
   (A color version of this figure is available in the online journal.)
   }\label{fig:TGamma}
\end{figure}

Except for two objects, the results from the case-by-case analyses of the 
spectral slope are less obvious. The low X-ray-to-\oiii\ luminosity 
ratio and their position on the MEx diagram suggests that these X-ray 
detected objects host X-ray absorbed AGN activity, which may appear 
to be in conflict with the steep X-ray spectral slopes found for the 
majority of these systems. However, the steep X-ray 
spectral slopes do not preclude the possibility that all of these
objects are heavily absorbed or Compton-thick AGN because the soft X-ray
emission could be dominated by either scattered nuclear emission or
star formation \citep[e.g.,][]{mat00}.  Indeed, known Compton-thick 
AGNs span a range of spectral slopes from very flat to very steep 
(Figure~\ref{fig:TGamma}), encompassing the range that we observe in our 
intermediate-redshift sample.  The nearby galaxies in Figure~\ref{fig:TGamma} 
are taken from \citet{bas99} and divided between Compton-thin and 
Compton-thick at $N_H=10^{24}$~cm$^{-2}$ (also see Figures~\ref{fig:OiiiXM} 
and \ref{fig:TLOiii}).  Overall, it is therefore possible that all of our
Compton-thick AGN candidates are genuinely absorbed by Compton-thick material (colored symbols on 
Fig.~\ref{fig:TGamma}); however, some of them 
cannot be unambiguously identified as such based on the X-ray slope alone. 
Higher energy observations ($>$10~keV) may help to confirm the presence 
of Compton-thick AGNs in galaxies with a steep X-ray slope.

\subsection{Linking AGN Absorption and Optical Classification}\label{sec:MEx_Abs}

\begin{figure}
\epsscale{1.} \plotone{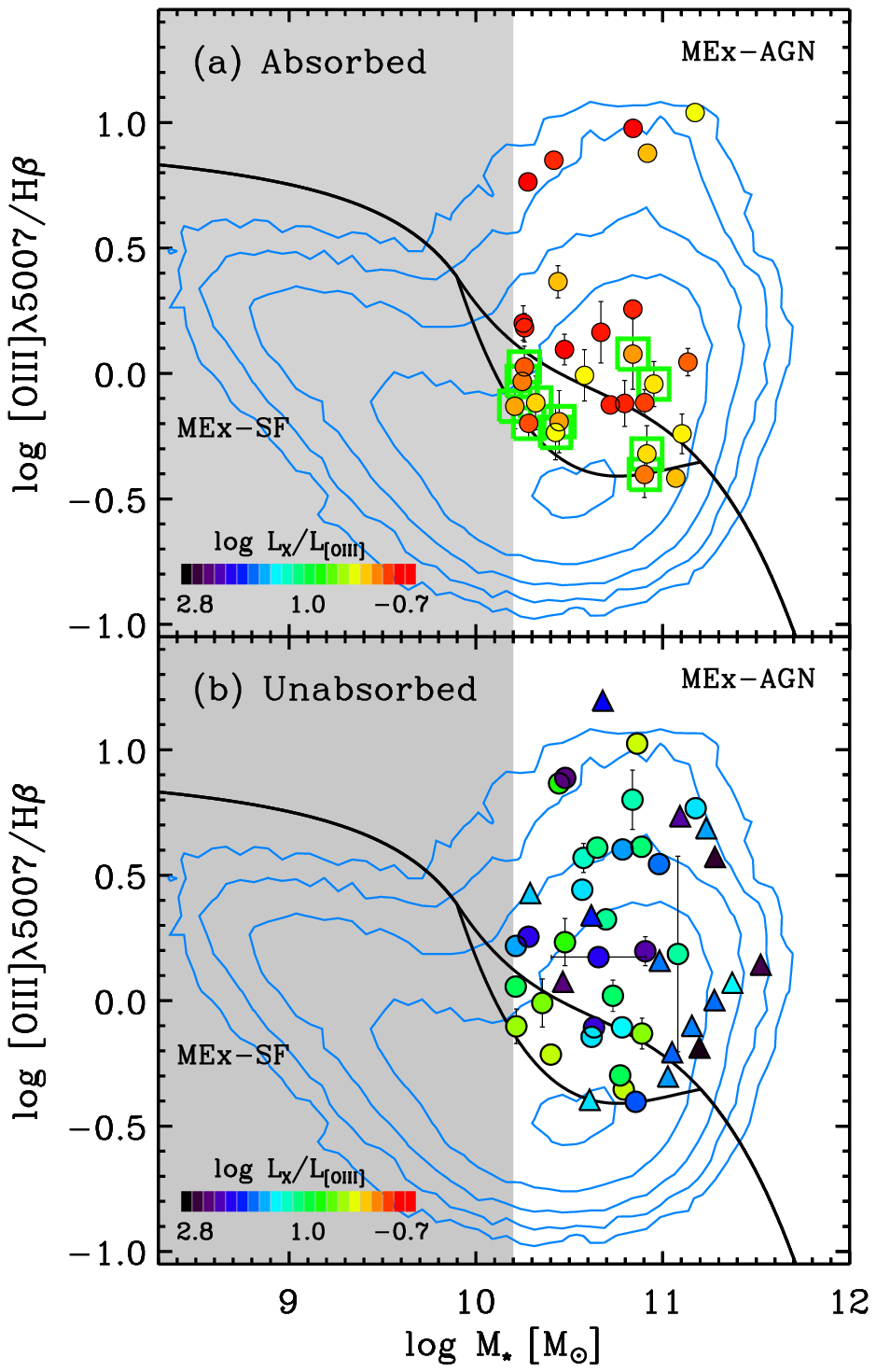}
\caption{
   Systems with P(AGN)$>$30\% and log($M_{\star}$[$M_{\sun}$])$>$10.2 shown on 
   the MEx diagram.  The points are color-coded according to the 
   Compton-thickness parameter $T$ shown in Figure~\ref{fig:TLOiii}.
   Red corresponds to the most absorbed systems, with log($T$)$\approx$-0.7 
   and the color changes gradually toward violet for the largest values of 
   log($T$) ($\approx$2.8).  Galaxies where both \hb\ and \oiiilam\ lines are 
   robust detections ($S/N>3$; filled circles) are shown in panels (a) and (b) depending if 
   they are absorbed [log($T$)$\leq$0.25] or largely unabsorbed [log($T$)$>$0.25].
   We also show galaxies with detected \oiiilam\ lines and \hb\ upper limits 
   (triangles in panel b).  In these cases, the \oiiilam/\hb\ ratio is a lower limit.   
   The green squares in panel (a) mark the X-ray undetected galaxies that yielded a 
   flat spectral slope when stacked.  Those tend to lie in the MEx-intermediate 
   region indicating that some galaxies with a buried AGN are also undergoing 
   an episode of active star formation.
   }\label{fig:mex_X}
\end{figure}

Next, we show the X-ray absorbed AGN candidates on the MEx diagram in 
Figure~\ref{fig:mex_X}(a).  Interestingly, their location differs from 
that of the X-ray unabsorbed systems [Figure~\ref{fig:mex_X}(b)].   
The latter are distributed evenly in the main MEx-AGN region 
with a possible bias toward high stellar mass.  In contrast, the 
absorbed AGN candidates cluster in the MEx-intermediate region with a few 
exceptions along the low-mass and outermost contour of the AGN plume.
As a reminder, the MEx-intermediate region is populated by BPT-composites 
and BPT-SF galaxies. This suggests that the X-ray absorbed AGN candidates 
are more likely composite systems than Seyferts or LINERs.

While we consider all objects (X-ray detections and non-detections) together, 
we highlight the subset of 13 galaxies that were not individually detected 
but yielded a very flat X-ray spectral slope of $\Gamma\sim0.4$  (see green 
squares on Fig.~\ref{fig:mex_X}).  The flat slope makes these systems robust 
Compton-thick AGN candidates. They occupy the MEx-intermediate region and 
the lower part of the AGN region where different AGN sub-classes overlap.  
The average probabilistic classification is P(AGN)$=58\%$ with the AGNs 
three times more likely to belong to the composite sub-class than the Seyfert~2 
and LINER sub-classes.  This result strengthens our conclusion that composite 
galaxies, hosting both star formation and AGN, are a very important population 
to search for highly-absorbed AGNs.  We further discuss these findings and 
their implications in \S\ref{sec:comp}.

\section{Discussion}\label{sec:discu}

Several previous studies have presented AGN diagnostic diagrams as 
alternatives to the BPT diagram.  Because the \niilam/\ha\ line ratio is 
only available in optical spectra out to redshift $\sim$ 0.4, it needs 
to be replaced in order for the diagnostic to be applicable at higher 
redshift while avoiding the need to obtain near-infrared spectra.
In this section, we first compare the MEx diagram with a few other AGN 
diagnostics from the literature (\S\ref{sec:compare}).  We then discuss how the MEx diagram 
contributes toward a more complete census of AGNs (\S\ref{sec:census}).  
We follow with a discussion of the importance of composite systems in this regard 
(\S\ref{sec:comp}) and we compare the fraction of absorbed AGNs from our work to 
results from the literature (\S\ref{sec:absfrac}). Finally, we mention 
possible evolutionary effects that are relevant to all nebular line 
diagnostic diagrams (\S\ref{sec:evol}).

\subsection{Comparison with existing diagnostics}\label{sec:compare}

\citet{lam04} developed an AGN diagnostic diagram that, like the MEx 
diagram, is designed to be applicable to optical spectra of galaxies 
out to $z \sim 1$.  In this case, the BPT diagram was modified by 
replacing the \niilam/\ha\ ratio with \oiilam/\hb.  Because of the greater 
wavelength separation between \oiilam\ and \hb, their flux ratio is very 
sensitive to dust obscuration.  Thus, the authors opted for an equivalent 
width ratio to mitigate against that effect.  The dividing lines on 
that diagram were recently revised \citep{lam10}.

The so-called blue diagram (because it includes only blue lines 
compared to \ha) has the advantage of splitting the LINERs from the 
Seyfert 2s.  However, it suffers from confusion between other classes 
of galaxies (SF, Sy2, and composites, see Lamareille et~al 2004; 
Lamareille 2010; and Appendix~C).  The MEx diagram has a notably 
cleaner separation between SF and Sy2 galaxies, is more sensitive to 
BPT-composite galaxies, and includes virtually all AGNs selected 
in the blue diagram.

In the case of the intermediate-redshift sample used here, a lot of galaxies 
with DEEP2 spectra have an insufficient wavelength coverage and do not contain 
all the lines required for the blue diagram.  As a consequence, the number of 
galaxies for which we can use this diagram is much smaller than for the
MEx diagram.  The modest wavelength range required for the spectra makes 
the latter more versatile than the blue diagram, which requires $\Delta\lambda_{rest} > 2900$\,\AA.

For the galaxies that have all the required observations, we find that 
the X-ray AGN selection has a much better agreement with the MEx diagram 
(Appendix~C).  This seems to be partially due to the mixing between BPT-SF 
and BPT-Sy2 galaxies in one of the regions of the blue diagram, and the 
mixing between BPT-SF and BPT-composites in another region.

The approach most similar to the MEx diagram was developed by \citet{wei07}; 
in fact, those authors used datasets that overlap with the present study.  
They combined TKRS and DEEP2 spectra in order to benefit from both the larger 
wavelength coverage of TKRS and the larger number of galaxies in the DEEP2 
survey.
They investigated the trends between \nii/\ha\ and \oiii/\hb\ as a function of 
absolute $H$-band magnitude ($M_H$), and found that \nii/\ha\ was sufficiently 
correlated with $M_H$ for the latter to be a helpful parameter when only the 
\oiii\ and \hb\ lines are available.  Weiner and collaborators did not go as
far as developing a diagnostic line, but showed a division between the red and 
blue galaxies on those plots in the sense that the red galaxies are predominantly 
occupying the region where AGNs are expected to lie.  This suggests that optical 
color may be another useful discriminant between star-forming and AGN galaxies 
when combined with \oiii/\hb.  Indeed, this approach was used recently by 
\citet{yan11}.

Yan and collaborators designed a similar diagram as presented here, except 
for using rest-frame $U-B$ color in lieu of stellar mass.  Their diagnostic 
produces comparable results to the MEx diagram when classifying the bulk of SDSS galaxies into 
AGN or star-forming.  In detail, they chose to avoid targeting the 
composite galaxies that lie beyond the distribution of the more obvious 
AGNs (i.e., above the Kewley line in the BPT diagram).  This implies that 
the galaxies on the star-forming side suffer from more frequent 
contamination by AGNs.  Thus, we identified the region where the 
composite galaxies have an important overlap with star-forming galaxies 
(region between Eq.~3 and Eq.~4).  The motivation is twofold.  First, it 
provides a means to obtain a cleaner sample of star-forming galaxies.  
Secondly, we have shown that composite galaxies are an important population 
to look for highly-absorbed AGNs that are co-existing with star formation 
in their host galaxies.  More detailed comparisons between the MEx 
diagram and the CEx diagram developed in Y11 are presented in 
\S\ref{sec:prob} and Appendix~\ref{app:class}.

Another alternative diagram, called the DEW diagram, was developed by 
\citet{sta06}.  It includes a separation based on the 4000\AA-break and 
equivalent width of \neiiilam\ or \oiilam\ (whichever is greater).  
In this case, the AGN hosts tend to have a larger $D_n(4000)$ with 
respect to normal star-forming galaxies.  Unfortunately, the 4000\AA-break 
is located near, or directly over, the detector gap for most of the 
DEEP2 spectra at the redshift of interest (peaking near $z=0.7$), 
so we are unable to apply this diagram for the bulk of our sample.  
The authors presented evidence that the DEW diagram has a better 
correspondence with the BPT diagram compared to the blue diagram 
(at least prior to its revision in 2010).

One notable difference between the MEx diagram and the original 
BPT diagram (or blue diagram) is that the former is not scale-free.  
The inclusion of stellar mass as a parameter imposes an absolute 
physical scale to the problem (whereas emission line ratios do not).  
This aspect raises questions about possible redshift evolution 
effects that could systematically affect the locus of the galaxies 
on diagnostic diagrams.  We discuss such possibilities below 
(\S\ref{sec:evol}).

\subsection{A More Complete Census of AGNs}\label{sec:census}

Understanding active galactic nuclei and their role in galaxy 
evolution requires a robust AGN classification scheme.  A complete 
sample should include both intrinsically weak and intrinsically 
bright but absorbed AGNs.  Weak AGNs are interesting to study and learn more 
about the low accretion phase and whether the AGN unified model needs 
to be revised \citep{ho08,tru09}.  Absorbed AGNs are sought to explain 
the unresolved portion of the cosmic X-ray background, and to 
quantify their contribution to mid- and far-infrared emission 
seen in infrared-luminous galaxies.  They are also of interest 
to test evolutionary scenarios where galaxy major mergers are 
invoked as a mechanism to form quasars, through a deeply-embedded
and absorbed phase \citep{san88,fab99,hop05}.

The MEx diagram introduced in this paper is a tool to uncover AGNs that 
are weak in X-ray observations, presumably due to either intrinsic 
weakness or X-ray absorption.  Both of these AGN phases -- 
weak or absorbed -- may be found in galaxies that are simultaneously 
undergoing episodes of star formation, potentially masking AGN 
signatures.  One important feature that allows us to find these 
systems is our probabilistic approach.

As shown in \S\ref{sec:Xstack}, using $P(AGN)>30\%$ as a threshold 
provides a more complete census of AGN than using $P(AGN)>50\%$.  
We found that both high-probability AGNs ($P(AGN)>50\%$) and 
medium-probability AGNs ($30\%<P(AGN)<50\%$) that are not 
individually detected in the {\it Chandra} data show signs of 
X-ray absorption when we stacked their X-ray emission (from the
resulting flat X-ray spectral slope; see Table~\ref{tab:X}).  The galaxies with 
$30<P(AGN)<50\%$ are likely composite systems (see \S\ref{sec:Xstack}), 
and occupy a region of the diagram that overlaps with other classes, 
making them more challenging to identify without the probability 
approach.  Thus, identifying those AGNs is a step toward a more 
complete census of active galactic nuclei.

Our finding that the X-ray undetected galaxies identified using the MEx diagram
are likely absorbed AGNs prompted us to make a 
more systematic search for X-ray absorbed systems.
We probed X-ray absorption using the Compton-thickness parameter 
$T$ as described in \S\ref{sec:Xstack_Cthik} (see references therein).
We again found a significant detection in {\it Chandra}'s hard band and
a still flatter spectral slope ($\Gamma\approx0.4$) for galaxies that 
were not detected individually.  This result supports the idea that 
the X-ray-to-\oiii\ luminosity radio does select absorbed AGNs as expected.  
We also note some galaxies with a steep spectral slope ($\Gamma>1$) and a 
low value of $T$. These systems are also consistent with X-ray absorption 
given that they are within the range of values of $\Gamma$ and $T$ spanned 
by known Compton-thick AGNs that are nearby and easier to study in 
more detail (see Fig.~\ref{fig:TGamma}).
In these cases, the soft X-ray emission may arise from a scattered AGN 
spectrum or from contamination by a superimposed starburst component.  

To summarize the absorbed AGN samples, we first found signs of X-ray 
absorption in MEx-AGNs that were not detected at X-ray wavelengths.  
These samples were selected with no {\it a priori} knowledge of 
X-ray absorption 
and the results are tabulated in \S\ref{sec:Xstack} (Table~\ref{tab:X}).   
We then specifically targeted X-ray absorbed AGN candidates by imposing 
$\log(L_{2-10keV}/L_{\oiiilam})<0.25$ (\S\ref{sec:Xstack_Cthik}).  
This yielded a sample of 33 absorbed AGN candidates: six have bright 
X-ray fluxes (and mostly a steep spectral slope, see red squares on 
Fig.~\ref{fig:TGamma}); ten have faint X-ray fluxes and the stacked 
signal is characterized by a slightly steep slope (blue shaded 
region on Fig.~\ref{fig:TGamma}), and of the remaining 17 that lack 
detections, 13 could be stacked and yielded a very flat spectral slope.  
So we have a sample of 33 candidates with heavy absorption 
of their X-ray emission, including a subsample of 15 more robust candidates 
(the 13 non-detections that were stacked plus the two individual detections 
that have a flat X-ray spectral slope).  

\subsection{Notes on Composite Galaxies}\label{sec:comp}

The separation between BPT-composites and BPT-AGNs is likely not as sharp as 
illustrated on the BPT diagram (Figure~\ref{fig:BPT}(a)) and may instead 
be a continuum of fractional AGN contribution to the spectral emission 
lines used in the diagnostic.  
This was suggested by, e.g., \citet{kew06}, who introduced a parameter to measure 
the distance from the star-forming sequence on the BPT diagram \citep[also see][]{yua10}.
The AGN branches defined in \citet{kew06} start from the metal-rich end of the 
star-forming sequence and follow mixing sequences toward larger values of \siilam/\ha\ or 
\oilam/\ha\ depending on the diagram used.  The idea is that there is a continuous 
transition from star-forming toward AGN with an increasing AGN contribution.

Like composites, which host both star formation and AGN, galaxies classified as
BPT-AGNs (above the Kewley line) may also have a significant star formation rate (SFR), 
up to $\sim 10\,M_{\sun}{\rm yr}^{-1}$ \citep{sal07}.  
On average their SFRs will be less than that in the population of composite 
galaxies \citep{sal07}.  One consequence is that the emission lines of the 
BPT-composites are characterized by a smaller fractional contribution from AGN 
relative to star formation, which makes the line ratios more ambiguous.  
These results also imply a continuity (or mixing sequence) between star-formation 
and AGN-dominated systems.  The so-called composites may be in the midst of a 
transition between the two types.

We might expect that some X-ray absorbed AGNs would reside in BPT-composite 
galaxies because large amounts of gas can provide fuel for star formation and 
act as absorbing material that attenuate X-ray signatures. In this case 
the X-ray absorption could come from the host galaxy's ISM in addition to 
torus-scale absorption.  In addition, if the gas is also mixed with dust, 
or if the galaxy is viewed through a dust lane, AGN emission-line 
signatures such as \oiiilam\ can be weakened and further diluted 
with emission from peripheral star formation \citep{mai95,mal98,gou09}, 
causing the global signal to exhibit composite signatures.

The degree to which the host galaxy ISM absorbs X-ray and \oiii\ 
emission may vary depending on the detailed geometry of the emitting 
and absorbing regions. The X-ray emission arises on the very small scales 
of the accretion disk, while \oiii\ is emitted on the larger scale of 
the narrow line regions (reaching several hundred parsecs to kiloparsecs 
from the nucleus).  If the absorption from the host galaxy ISM were 
approximated as a uniform screen with an extent that covers both the 
X-ray and narrow line emitting regions and with a Milky Way-like 
dust-to-gas ratio, then we would expect optical AGN signatures (such 
as \oiiilam) to be obscured more than the X-ray emission.  An extinction 
of $A_V=10$~mag would fully obscure \oiii\ emission from the narrow line 
regions, but would correspond to a gas column density of 
$N_H=1.6\times10^{22}$~cm$^{-2}$, causing only modest absorption of the 
X-ray emission.  This would suppress the $L_{\oiii}/L_{2-10keV}$ ratio, 
leading to the opposite effect of the preferential X-ray absorption that 
we observe in some systems.  

However, if the foreground ISM is clumpy, then it is possible for a dense 
cloud with large $N_H$ to obscure the compact X-ray emitting region, 
while the average absorption to the larger \oiii-emitting narrow line 
regions could be much less.  There will be lines of sight to the 
extended narrow line regions with more modest obscuration which will dominate 
the global emission line signal (relative to more heavily obscured lines of sight).  
As a result, we may in this case expect a preferential absorption of the 
X-ray emission relative to the optical (\oiiilam) emission.

The preferential extinction of the small-scale X-ray emission when compared 
to the larger scale \oiii\ emission could also be caused by a physical 
connection between the gas at galaxy scales and the gas on the scales of 
the AGN torus.  For example, in the simulations of 
\citet{hop10}, an overall larger gas fraction on kpc scale can result in 
more gas funneling to the inner sub-pc region of the systems.  
These simulations predict that the transport of gas toward the central 
region depends most strongly on the disk-to-bulge ratio 
and on the gas fractions on $100-300$~pc scales.  We note that these authors 
did not specifically model a torus.  However, the presence of instabilities 
on small scales increases the gas flow to the inner regions and suggests that the 
torus, often thought to be an extension of the accretion disk, could contain 
more obscuring material (or have a larger filling factor). 
If this scenario were true, it would imply that a higher gas fraction on galaxy 
scales could lead to both a larger SFR and a more gas-rich accretion disk 
(and potentially torus), thus creating preferential absorption of small scale 
emission (hard X-rays) relative to larger scale emission (\oiiilam).

Based on these scenarios, and on our observations of a hard X-ray signal by stacking 
only 12 likely composite galaxies (with $30<P(AGN)<50$\%, \S\ref{sec:Xstack_Cthik}) that were 
not detected with very sensitive {\it Chandra} observations, we propose that 
not all BPT-composite galaxies have intrinsically weak {\it transition} AGNs.  Instead, a 
fraction of them have powerful but absorbed AGNs whose light is diluted with that 
of their host galaxies (regardless of whether the host galaxy ISM provides additional 
 AGN absorption or not).

In addition, the composite galaxy populations may differ with redshift.  In general, 
star-forming galaxies had a higher specific star formation rate ($SSFR \equiv SFR/M_{\star}$) 
at earlier times \citep{noe07,elb07}.  The larger amount of star formation at higher redhsift 
is also linked to larger reservoirs of molecular gas \citep[e.g.,][]{dad10,tac10}.  
As mentioned before, large amounts of gas contribute to increase the column density along 
the line of sight thus X-ray absorption.  The more sizable gas reservoirs in isolated 
galaxies at higher redshift may be a way to obtain more absorbed AGNs without major 
mergers.  

There is also evidence for a larger infrared-to-X-ray 
luminosity ratio with increasing redshift \citep{mull10}, further suggesting a larger 
SSFR in the host galaxies of higher-redshift AGNs relative to the current epoch. The 
enhanced star formation may affect the line ratios in the sense that more galaxies 
will be classified as BPT-composites at higher redshift.  While we do not study star 
formation rates in this work, we remind the reader that the composite galaxy population 
plays an important role in the search for X-ray absorbed AGNs and as such should be 
identifiable at high-redshift.  Thus, the MEx diagram and the probabilistic classification 
scheme introduced in this Paper are expected to fulfill this need and to contribute to a 
significant improvement in identifying the population of missing Compton-thick AGNs.

\subsection{Absorbed AGN Fractions}\label{sec:absfrac}

Next, we examine the absorbed AGN fraction in terms of the
number of galaxies identified as well as fractional contribution to $L_{\oiii}$.
Galaxies with log($L_X/L_{\oiii}$)$>$1 are considered as unabsorbed
while galaxies with $\log(L_X/L_{\oiii})<0.25$ are considered absorbed and
likely Compton-thick and galaxies in between are likely absorbed but 
Compton-thin with $22<\log(N_H[{\rm cm^{-2}}])<24$ [Figure~\ref{fig:OiiiXM}(a)].

For galaxies in EGS, we compute X-ray upper limits using sensitivity and exposure time 
maps\footnote{Data products from http://astro.ic.ac.uk/content/chandra-data-products}.  
We convert the count rates in the hard band to fluxes assuming the conversion 
factor used by \citet{lai09}.  The X-ray detection limit is highly variable across the 
EGS field.  For the \oiii-selected samples that we consider here, it varies from 
$6.6 \times 10^{-16}$ to $1.2 \times 10^{-14}$\,erg\,s$^{-1}$\,cm$^{-2}$.  
The average value is $2.5 \times 10^{-15}$\,erg\,s$^{-1}$, similar to the flux 
limit at which the survey is complete over 50\% of the area 
\citep[$2 \times 10^{-15}$\,erg\,s$^{-1}$\,cm$^{-2}$;][]{lai09}.  
We discard the X-ray upper limits of 10\% (7/70) of \oiii-selected galaxies with 
$L_{\oiiilam}>10^{41}$\,erg\,s$^{-1}$ because they lie in very shallow regions of the 
X-ray data (with upper limits that are $> 5 \times 10^{-15}$\,erg\,s$^{-1}$\,cm$^{-2}$).  

However, we note that a typical upper limit value of $2.5 \times 10^{-15}$\,erg\,s$^{-1}$\,cm$^{-2}$  
does not constrain the absorption very tightly, especially at 
log($L_{\oiii}$[erg~s$^{-1}$])$<$41.5 where a lot of upper limits correspond 
to log($T$)$>$1.  Thus, we derive lower and upper limits to the absorbed AGN 
fraction by assuming that these galaxies are respectively all unabsorbed 
(i.e., they actually lie at log($T$)$>$1) or all absorbed (i.e., they lie 
at log($T$)$<$1).  We furthermore weigh the galaxies with the probability that 
they host an AGN according to the MEx diagnostic.  The fractional number and 
\oiii\ luminosity contribution of absorbed AGNs are respectively defined as follows:  
\begin{eqnarray}
 f_{absorbed} &=& \frac{\sum_{i=1}^{N_{absorbed}} P(AGN)_{i}}
                     {\sum_{j=1}^{N_{total}} P(AGN)_{j}}   \\
 f(\oiii)_{absorbed} &=& \frac{\sum_{i=1}^{N_{absorbed}} P(AGN)_{i} \times L(\oiii)_i} 
                     {\sum_{j=1}^{N_{total}} P(AGN)_{j} \times L(\oiii)_j} \label{eq:Oiiifrac}
\end{eqnarray}
where $P(AGN)$ is the fractional probability of hosting an AGN varying from 0 to 1, 
$N_{absorbed}$ is the number of X-ray absorbed AGNs, and $N_{total}$ is the total number 
of AGNs (absorbed and unabsorbed).  
At log($L_{\oiii}$[erg~s$^{-1}$])$>$41, the absorbed (Compton-thick) fractions are 
poorly constrained: we find $f_{absorbed}=25-81$\% ($f_{Compton-thick}=12-81$\%).  
Using equation~\ref{eq:Oiiifrac}, the fractional contribution of absorbed (Compton-thick) 
AGNs to the \oiii\ luminosity of all AGNs with log($L_{\oiii}$[erg~s$^{-1}$])$>$41 is 
$f(\oiii)_{absorbed} =$54$-$82\% ($31-82$\%).

The constraints are slightly better with a higher \oiii\ luminosity threshold.  
Restricting our analysis to the 34 galaxies at $0.5<z<0.8$ with 
log($L_{\oiii}$[erg~s$^{-1}$])$>$41.5, we find an absorbed (Compton-thick) fraction 
ranging from 45\% to 68\% (17\% to 68\%) in number of galaxies, and a contribution 
to $L_{\oiii}$ between 70$-$81\% (39$-$81\%).  Similarly, Y11 estimated an absorbed 
AGN fraction by calculating the X-ray detection probability assuming a constant 
intrinsic X-ray/\oiiilam\ ratio and measured \oiii\ luminosities, and by comparing the 
expected number of X-ray detections to the actual number of detections 
at the EGS depth.  Attributing the non-detections to X-ray absorption, they find  
an X-ray absorbed fraction of $\sim$50$-$60\% at the same \oiii\ luminosity threshold used 
here, in agreement with our estimate for EGS and GOODS-N combined ($45-68$\%).

Using only GOODS-N galaxies provides stronger constraints on 
X-ray absorption given the high sensitivity of the {\it Chandra}
observations in that field, but is subject to smaller number statistics.
In this case, restricting our analysis to the 29 galaxies at $0.3<z<0.8$ with 
log($L_{\oiii}$[erg~s$^{-1}$])$>$40.5 yields an absorbed (Compton-thick) 
fraction of 75$\pm$18\% (54$-$64\%).
For log($L_{\oiii}$[erg~s$^{-1}$])$>$40.5, Y11 calculated that 70.5\% of AGN 
fail to be detected at the EGS depth, suggesting an absorbed fraction 
($70.5\pm4.1$\%) that agrees with our calculations for GOODS-N galaxies (75\%).  
Our results are also consistent with \citet{aky09}, who studied 38 nearby ($<$70~Mpc) 
Seyferts and found an absorbed fraction 55$\pm$12\%.  
Considering only sources with $L_{2-10keV}>10^{41}$erg~s$^{-1}$, they find 
that 75$\pm$19\% of 21 galaxies are absorbed ($N_H>10^{22}$\,cm$^{-2}$) 
and 15$-$20\% are Compton-thick.

At quasar-like luminosities, \citet{vig10} found that 68\% of SDSS Type 2
quasars at $0.3<z<0.8$ are Compton-thick according to \oiii\ versus 
X-ray criteria.  As they discuss, an \oiii-selection may bias the numbers 
toward a larger absorption fraction. Our estimates do not allow us 
to predict whether the absorbed AGN fraction rises or declines with 
\oiii\ luminosity.  Larger samples with very sensitive X-ray observations 
would help to address that question.

\subsection{Possible evolutionary effects}\label{sec:evol}

As noted in \S\ref{sec:hiz}, there is some evidence that the 
intermediate-redshift galaxies may be offset in the MEx diagram relative 
to the lower-redshift SDSS sample.  Other authors have reported similar 
shifts in the BPT diagram, i.e., higher-$z$ galaxies appear displaced 
toward larger values of \oiii/\hb\ and/or \nii/\ha.  Some attributed 
this trend to varying HII region conditions \citep{liu08,bri08,hai09} 
while others claim that additional AGN contribution may be the driving 
factor \citep{gro06,wri10}.

In particular, \citet{wri10} presented a detailed study of one BPT-composite 
galaxy at $z\sim1.6$.  Using adaptive optics combined with integrated 
field spectroscopy, they were able to separate the AGN 
from the host galaxy emission and found that the central region (inner 
$0.\arcsec 2 \times 0.\arcsec 2$) occupies the AGN part of the BPT diagram, 
while the integrated measurements place this galaxy in the BPT-composite 
region and the host galaxy alone shares the locus of the normal 
star-forming galaxies (i.e., not shifted from the low-$z$ sequence).  
This suggests that diluted AGN contribution may explain the offset in 
at least some cases.  

What would be the consequences of such offsets on the MEx diagram?
If more galaxies have diluted AGN contributions, they will be selected 
as MEx-AGNs because the offset would tend to move the galaxies 
over the dividing line.  A fraction of the higher-mass
galaxies are already subject to be BPT-composite ($\sim 20-30$\%) 
so in this case, moving them over the AGN line would actually increase the  
completeness.  On the other hand, if galaxies had different physical
conditions in their HII regions in the past, in the sense of having 
larger \oiii/\hb\ ratios compared to galaxies today, 
then some purely star-forming galaxies could be moved over the AGN 
dividing line.  However, the differences out to $z \sim 1$ are 
probably small since the empirical offset is around 0.2~dex in log(\oiii/\hb).

Given that the situation is still under debate, 
and that selection effects are not well constrained, we do not 
implement an offset at this point.  However, we note that the very 
good agreement between the X-ray and the MEx selection even out to 
redshift $\sim 1$ supports the conclusion that the empirical division 
works well over the redshift range considered here.  Future work 
would be required to test whether it holds to even higher redshift.

\section{Summary}\label{summ}

In this paper, we provide a new tool to gather a more complete census of 
actively accreting black holes in galaxies.  We argue that intrinsically 
weak AGNs, as well as heavily absorbed AGNs, are important to understand 
the connection between supermassive black holes and their host galaxies.  
We successfully find systems that belong to these observationally-challenging 
AGN classes.

Using 110,205 emission-line galaxies from the SDSS, we calibrate a new 
excitation diagram to identify galaxy nuclear activity.  By combining 
the observed \oiiilam/\hb\ emission line ratio with stellar mass, we 
obtain a diagnostic applicable to galaxies out to $z \sim 1$.
Here we summarize a few properties of the Mass-Excitation (MEx) diagram:

\begin{enumerate}

\item The simplest version splits galaxies into three classes: MEx-AGN, 
  MEx-SF and MEx-intermediate.  The latter is found in a region where 
  BPT-composite galaxies have significant overlap with star-forming galaxies. 
  The classification has an excellent correspondence to that from the BPT 
  diagram based on \oiiilam/\hb\ and \niilam/\ha.

\item The classification scheme is refined using a novel probabilistic 
  approach to predict the location of high-redshift galaxies on standard 
  BPT diagrams based on their \oiiilam/\hb\ ratio and stellar mass (\S\ref{sec:prob}).
  Utilizing SDSS priors, we calculate the probability of four mutually 
  exclusive spectral 
  classes (star-forming, composite, LINER, or Seyfert~2) as a function of 
  position on the MEx diagram for each galaxy.

\item The MEx diagram offers a more complete AGN selection than 
  alternatives such as the blue diagram \citep{lam10}.  It is comparable 
  to the Color-Excitation (CEx) diagram developed by \citet{yan11} but 
  may be particularly useful for dusty, strongly reddened galaxies which 
  may have unusual colors.  Furthermore, we add a dividing line to 
  the CEx diagram to identify composite galaxies, an important population 
  to search for absorbed AGNs.

\item We provide publicly available IDL routines to calculate the 
  probabilities of different galaxy spectral classes based on the MEx 
  and CEx diagrams.

\end{enumerate}

Another outcome of a nearly complete AGN selection is that the complementary  
selection results in a very clean star-forming galaxy sample.  Indeed, 
with our probability scheme, one can weight galaxies as a function of P(SF) 
when computing average properties such as the stellar mass-metallicity 
relation or the global star formation rate and thus mitigate against AGN 
contamination.  We anticipate that a wide variety of applications will 
benefit from our approach to galaxy spectral classification.

We successfully classify AGN and star forming galaxies at $0.3 < z < 1$ 
using our new diagnostic. The sample is drawn from the AEGIS and GOODS-N 
surveys. Our main findings are as follows:

\begin{itemize}
\item An independent X-ray classification scheme shows that the MEx diagram 
  selects around 82\% of the X-ray AGNs with detected emission lines (S/N$>$3
  for \hb\ and \oiiilam).  
  When considering all X-ray AGNs, we find that 59\% are 
  in the MEx-AGN or MEx-intermediate regions, 8\% are in the star-forming 
  region, and 33\% lack a classification (they fail the $S/N>3$ requirement 
  for valid emission line detections).

\item The MEx diagram reveals X-ray absorbed AGN candidates, 
  which were missed or mis-identified in X-rays while having robust signatures 
  in their optical spectra.  We combine support from three lines of evidence:
  \begin{enumerate}

  \item[1.] The subset of these galaxies with a low enough redshift to be 
    placed on the BPT diagram lie in the BPT-composite or BPT-AGN regions 
    (Fig.~\ref{fig:bptX}).  In addition, one galaxy has 
    a clear detection of \nevlam\ in emission, a secure AGN tracer 
    (\S\ref{sec:MEx_X}).

  \item[2.] Stacking {\it Chandra} data of galaxies that were not 
    individually detected in GOODS-N reveals a hard signal and a 
    flat X-ray spectral slope for galaxies that have $P(AGN)>30\%$.  
    This suggests that at least some of the 32 galaxies that were 
    stacked do host an actively accreting black hole (\S\ref{sec:Xstack}).  

  \item[3.] By combining \oiiilam\ and hard X-ray luminosities, we calculate a 
    Compton-thickness parameter ($T \equiv L_{2-10keV}/L_{\oiiilam}$).
    We identify 33 highly-absorbed AGN candidates.  The presence of AGN 
    in at least some of the candidates is supported by the hard X-ray 
    signal ($\Gamma \sim 0.4$) obtained from stacking 13 galaxies that 
    were not detected individually (\S\ref{sec:Xstack_Cthik}).  
    We note that some X-ray galaxies have a steep spectral slope despite
    a low value of $T$.  These objects are also consistent with the range
    of values expected from a comparison with known, nearby Compton-thick
    AGNs (Fig.~\ref{fig:TGamma}). 
  \end{enumerate}

\item Many absorbed AGNs in the intermediate redshift sample are composite systems, 
  with both star formation and active black hole accretion. The increase in the 
  SF:AGN ratio -- and hence a decrease in \oiiilam/\hb\ -- may be related 
  to the global rise in the specific star formation rate with redshift.  However, 
  the details of such a connection remain open questions (does the host galaxy ISM 
  provide additional AGN absorption? is there a link between host galaxy and torus 
  properties?).  The identification of these composites poses a considerable challenge 
  on the observational point-of-view.  The MEx probabilistic approach proved to be key 
  in this respect.

\end{itemize}

On-going and future near-infrared multi-object spectroscopy 
surveys will shed more light by allowing us to observe all the 
traditional emission lines directly.  The MEx diagram will nevertheless 
be useful to fill in gaps where \ha\ and/or \niilam\ fall in especially 
noisy regions of the spectra or where \niilam\ fails to be detected 
(as it is typically fainter than \ha).  In principle, the MEx diagram 
could be applied out to $z \sim 4$ using near-infrared spectroscopy, 
thus pushing the limits beyond what will be achievable when 
relying on \ha\ and \niilam\ detections (since the latter redshift 
out of the $K$-band at $z \sim 2.5$).

\acknowledgments We gratefully acknowledge the anonymous referee for 
useful suggestions that improved this manuscript.  The authors also 
thank C. Tremonti for kindly providing spectra stellar population fitting 
routines, M. Cooper for valuable help with the DEEP2 spectra, 
and E. Daddi for sharing an extended version of his GOODS-North 
photometric catalog.  SJ was partially funded by a FQRNT fellowship  
(Fonds Qu\'eb\'ecois de recherche sur la nature et la technologie, 
Qu\'ebec, Canada) and a Philanthropic Educational Organization Scholar Award.  
DMA gratefully acknowledges support from a Royal Society University
Research Fellowship and a Philip Leverhulme prize.

The authors would like to thank the many members of the GOODS team who 
obtained, reduced, and cataloged some of the data used in this paper.  
This work is based in part on observations made with the {\it Spitzer} 
Space Telescope, which is operated by the Jet Propulsion Laboratory, 
California Institute of Technology under a contract with NASA. Support 
for this work was provided by NASA through an award issued by JPL/Caltech.

Funding for the DEEP2 survey has been provided by NSF grants 
AST95-09298, AST-0071048, AST-0071198, AST-0507428, and AST-0507483 
as well as NASA LTSA grant NNG04GC89G. The analysis pipeline used to 
reduce the DEIMOS data was developed at UC Berkeley with support 
from NSF grant AST-0071048.

Some of the data presented herein were obtained at the W. M. Keck 
Observatory, which is operated as a scientific partnership among 
the California Institute of Technology, the University of California 
and the National Aeronautics and Space Administration. The Observatory 
was made possible by the generous financial support of the W. M. Keck 
Foundation. The Keck Observatory acknowledges the very significant 
cultural role and reverence that the summit of Mauna Kea has always 
had within the indigenous Hawaiian community and appreciate the 
opportunity to conduct observations from this mountain.

\appendix
\section{Comparing the MEx and CEx Diagrams}\label{app:class}

The mass-excitation (MEx) and color-excitation (CEx) diagrams are two useful 
alternatives when the emission lines typically used for AGN diagnostics
such as the BPT diagram are not available.  The former is developed in this paper 
(\S\ref{sec:bpt}) and involves using stellar mass as a discriminant between predominantly 
star-forming galaxies and those hosting an AGN.  The CEx diagram was developed by Yan et al.
(submitted) and involves using rest-frame $U-B$ color rather than stellar mass.
Galaxies hosting an AGN tend to be both massive and redder in $U-B$ so it is not surprising
that both these approaches give similar results for many galaxies.  

\begin{figure*}[thb]
\epsscale{1.} \plotone{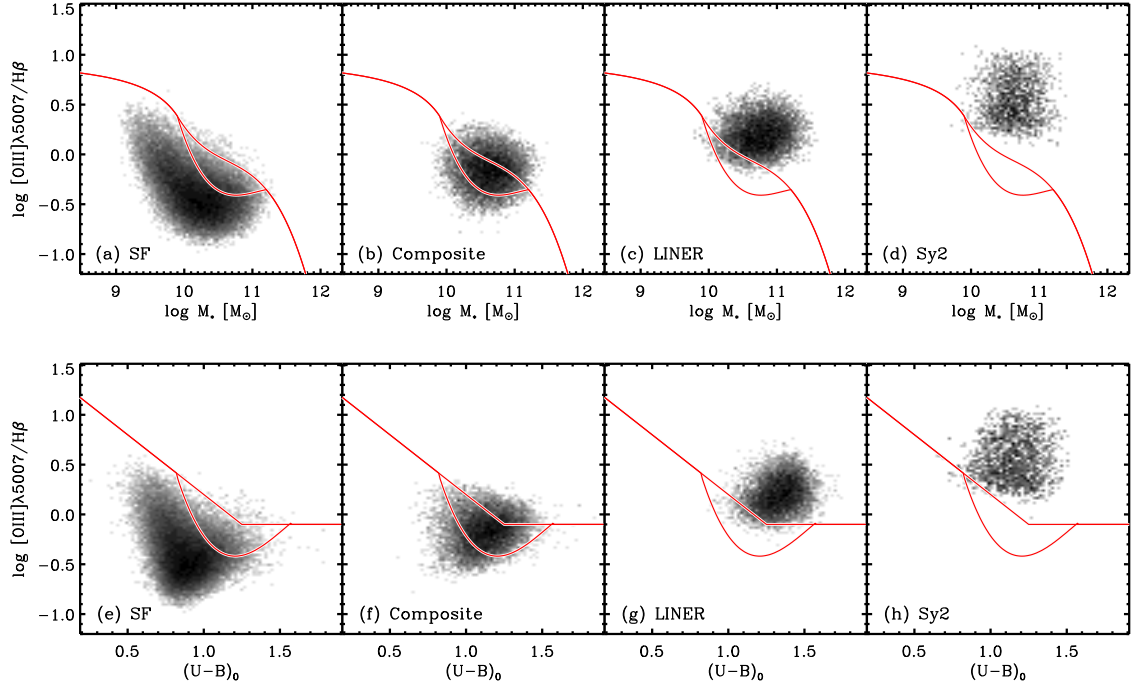}
\caption{
   Bivariate distributions of the SDSS galaxies on the MEx diagram (top row) and 
   CEx diagram developed by Y11 (bottom row).  Galaxies are plotted separately for
   each classification: (a) star-forming, (b) composite, (c) LINER, and (d)
   Seyfert 2.  The same order is followed in panels (e)-(h) for the CEx diagram.
   The lower 
   dividing lines were added to mark the region with significant overlap between 
   BPT star-forming and BPT-composite galaxies (the fraction of composites is 
   greater than 40$-$50\% between the lines).
   (A color version of this figure is available in the online journal.)
   }\label{fig:distr}
\end{figure*}

We compared the fraction of galaxies hosting AGN to the purely star-forming galaxies 
according to the BPT diagram in \S\ref{sec:prob}.  We defined an additional line on the 
CEx diagram to provide a means to identify a region where star-forming galaxies are mixed
with composite galaxies, analogous to our method with the MEx diagram.  Here we look at 
the bivariate distributions of galaxies split into the four categories defined in 
\S\ref{sec:prob}. Figure~\ref{fig:distr} shows that both diagrams are generally comparable.
The main dividing lines (upper lines) make a division between the LINERs and Seyfert 2's 
on the upper side and the star-forming galaxies on the lower side.  By design, the MEx
diagram selects a larger fraction of the composites in the AGN side.  This decision was
motivated by our goal to provide an increased completeness of AGN candidates including 
intrinsically weak as well as absorbed systems that may be undergoing star formation 
simultaneously with an AGN phase.  
We have shown the strong potential to find X-ray absorbed candidates among such a population 
of galaxies (\S\S\ref{sec:Xstack},\ref{sec:AGNpower}).

On the CEx diagram, we note a few galaxies in the star-forming category 
that seem to be outliers with very red $U-B$ colors (also see 
Figure~\ref{fig:prob}).  These may be especially dusty.  Their number 
is small relative to the bulk of the star-forming population in the 
SDSS sample but they may be interesting targets in a different context.
For instance, we may expect more star formation and active galactic 
nuclei to take place in dusty, infrared-luminous galaxies at higher 
redshift \citep{lef05,mag09}.

\clearpage

\section{Stellar Mass Estimation}\label{app:mass}

There are 3174 galaxies with both a stellar mass estimate from SED 
fitting and absolute rest-frame $K$-band magnitude from observed
IRAC photometry.  Here, we use EGS galaxies with a SED fit to the
FUV, NUV, $ugriz$, and $K$ photometry and GOODS-N galaxies with a 
SED fit to $UBVRIzJK$ bands (\S\ref{sec:mass}).

\begin{figure*}[bh]
\epsscale{0.75} \plotone{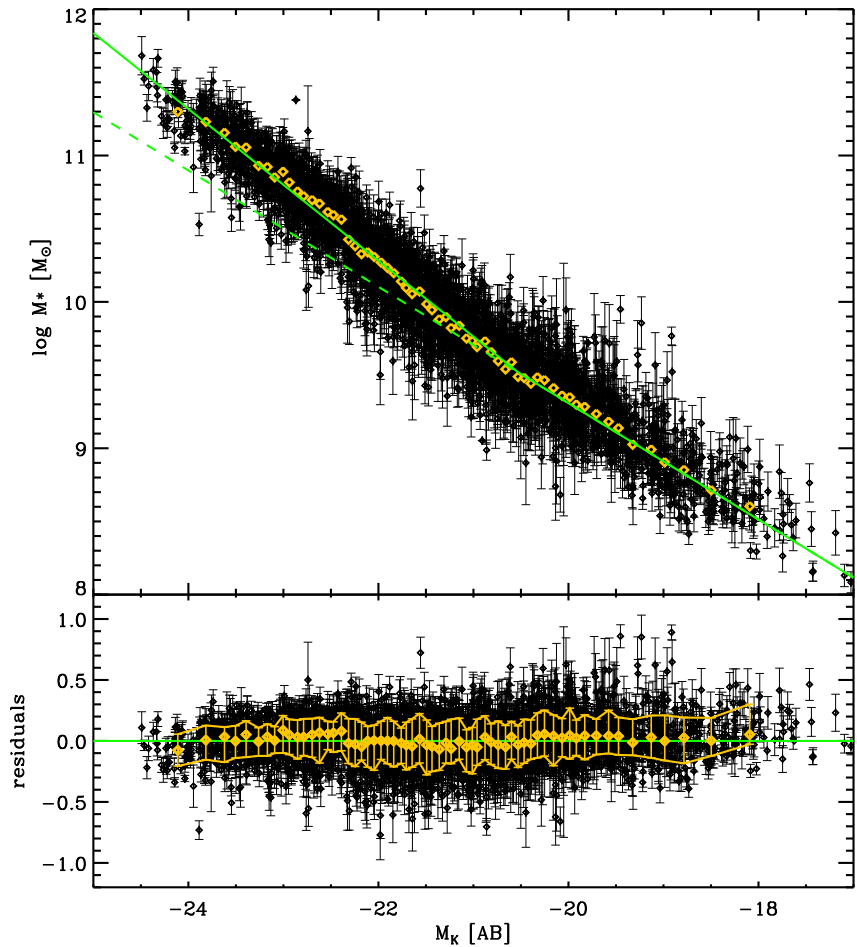}
\caption{
   [Top] Stellar mass obtained from SED fitting as a function of absolute 
   rest-frame $K$-band magnitude ($M_K$, in AB magnitudes).  We fit two 
   mass-luminosity power laws for the bright and faint ends, which intersect 
   at $M_K = -20.5$~AB (green lines). The dashed line is the extension of the fit
   at the faint end, with a linear slope between mass and luminosity.  For reference, 
   the yellow diamonds show the median log($M_{\star}$) in $M_K$ bins (every 51 points).  
   [Bottom] 
   Residuals of the top panel as a function of $M_K$.  Yellow diamonds and error 
   bars show the median and the 16th to 84th percentile range.  There is no 
   obvious trend in the residuals (median=0.009~dex), and the overall dispersion 
   ($\approx (84PL - 16PL)/2$) is 0.18~dex.
   (A color version of this figure is available in the online journal.)
   }\label{fig:mass}
\end{figure*}

The relation between stellar masses derived from SED fitting and 
rest-frame K-band absolute magnitudes is displayed in Figure~\ref{fig:mass}.  
The broken power-law relation is expressed as:
\begin{align}
\log M_{\star} &= -0.398 \times M_K + 1.357;~~~~~for~M_K > -20.5~AB  \\
              &= -0.519 \times M_K - 1.128;~~~~~for~M_K < -20.5~AB
\end{align}\label{eq:mass}

Writing the relation as $M_{\star} \propto L_{K}^{\alpha}$, the slopes 
found in Eq. B1 \& B2 imply power-law indices $\alpha = 0.99$ at 
$M_K > -20.5~AB$ and $\alpha = 1.3$ at $M_K < -20.5~AB$.  This relation 
is linear at the faint end but slightly steeper at the bright end.  
The steepening of the slope is presumably due to the fact that more massive galaxies 
are redder owing to the older average age of their stellar populations, 
and thus have higher mass-to-light ratios.

The residuals of the broken power-law fit (bottom of Figure~\ref{fig:mass}) 
have a constant scatter with luminosity.  The dispersion of the overall sample 
is 0.18~dex, around the mean (median) of 0.004 (0.009)~dex.

Errors on the stellar masses calculated using Eq. B1 \& B2
are estimated by combining the average error on log($M_{\star}$) from galaxies 
with SED fitting (0.12~dex) and the dispersion on the log($M_{\star}$) $- M_K$ 
residuals (0.18~dex).  Adding these two contributions in quadrature yields 
0.22~dex, which we use for all galaxies lacking a SED fit.

\clearpage
\section{Comparison with the {\it Blue} Diagram}\label{sec:blue}

The blue diagram discussed in \S\ref{sec:compare} was developed in 
\citet{lam04} and recently improved in \citet{lam10}.  The diagnostic 
employs lines at blue rest-frame wavelengths, from \oiilam\ to \oiiilam, 
in order to facilitate its use out to $z \sim 1$.  This motivation is 
similar to that which guides the design of our MEx diagram, although our results differ significantly.  

In this diagnostic, the abscissa is an equivalent width ratio because the spectral separation
between the \oiilam\ and \hb\ lines is significant and their line flux ratio would thus
be very sensitive to dust attenuation.
Note the inability to distinguish between SF and Seyfert 2 galaxies in the region marked as SF/Sy2.
We show that this ambiguity is completely removed with our new diagram.

\begin{figure*}[bh]
\epsscale{0.8} \plotone{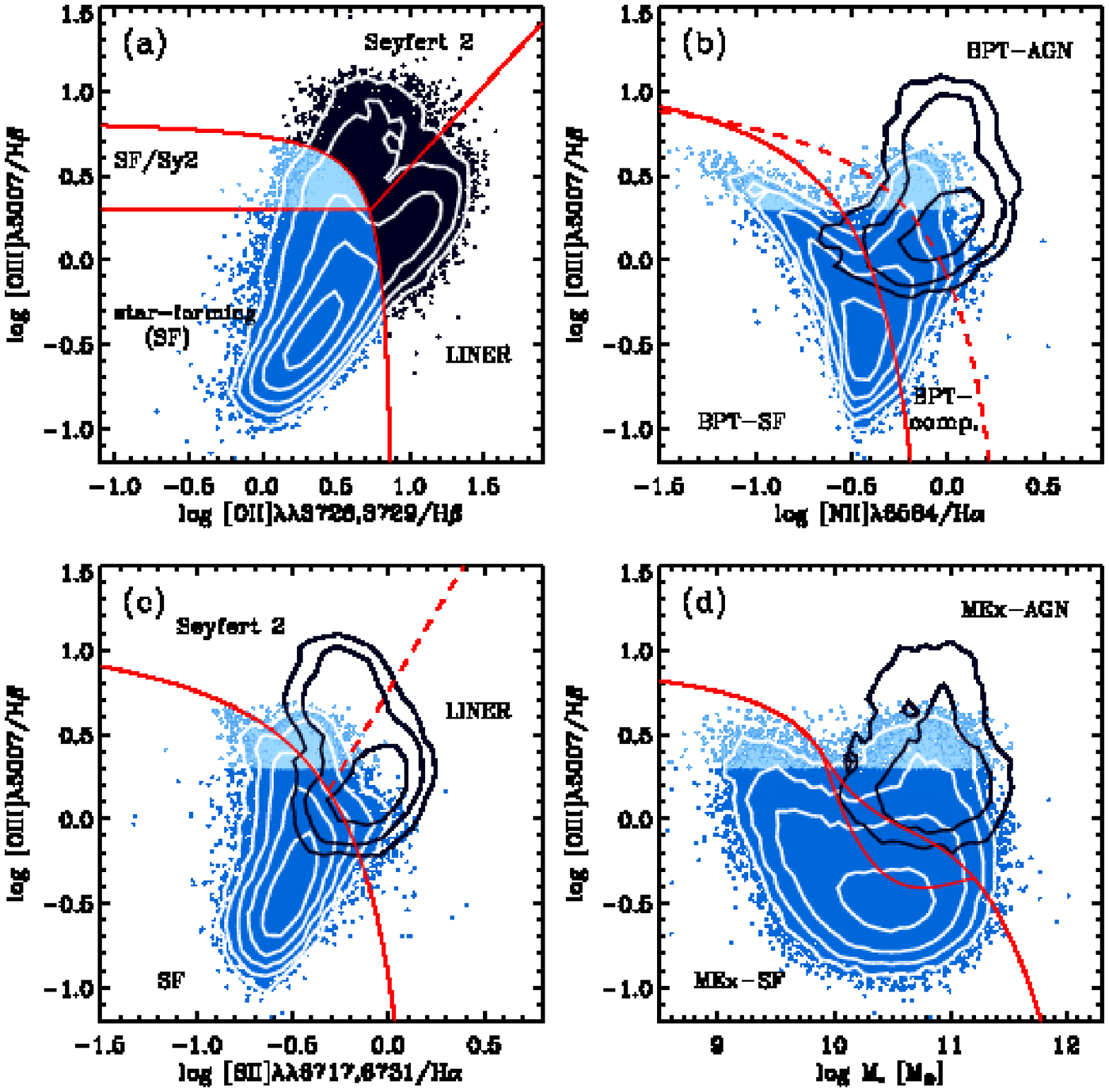}
\caption{
   Distribution of SDSS galaxies on the blue diagram: \oiiilam/\hb\ line flux ratio against 
   \oiilam/\hb\ equivalent width ratio.  This diagram and dividing lines are adapted 
   from \citet{lam10}.  Most star-forming 
   galaxies (SF) occupy the bottom left of the plot (blue dots) but include a number of composite 
   galaxies.  The black dots correspond to AGNs (Seyfert 2s and LINERs), and the light blue
   points mark the galaxies in the region where SF and Sy2 classes overlap.  These classifications
   are compared to other diagnostics: (b) \nii\ BPT diagram, (c) \sii\ BPT diagram, and (d) MEx diagram.
   The blue diagram star-forming galaxies include a number of composite galaxies as shown 
   on the BPT diagram (b), as well as AGNs that are mostly LINERs according to panel (c).  
   The ambiguous SF/Sy2 galaxies (light blue) are well separated on the other panels, 
   especially (b) and (d). In all panels, the contours indicate the density of points 
   (in bins of 0.075 dex $\times$ 0.075 dex) and are logarithmic 
   (0.5~dex apart, with the outermost contour set to 10 galaxies per bin).
   (A color version of this figure is available in the online journal.)
\label{fig:blue}}
\end{figure*}

We examine the selection functions built in the blue diagram against the three 
other diagnostics introduced earlier. On Figure~\ref{fig:blue}, the blue points represent the 
star-forming galaxies \citep[panel (a); note that there is a known overlap with the composite 
galaxy population as discussed in][]{lam10}.  The pale blue points mark the ambiguous region
where SF and Seyfert 2 galaxies are indistinguishable (SF/Sy2) whereas the black dots indicate the 
AGNs (encompassing both Seyfert 2's and LINERs).  Panels (b)-(d) show the location of the same galaxies
over the other diagnostic diagrams.  The main features are that {\bf (i)} the ambiguous SF/Sy2 region (pale blue points) 
is resolved in the other diagrams, which all break the degeneracy observed in the \oiilam/\hb\ EW ratio
of low-$M_{\star}$ SF galaxies and Seyfert 2s; 
{\bf (ii)} the blue diagram AGNs (black dots and black contours) include some BPT-composites: 
more so than the \sii-diagram shown in (c), but less so than the BPT (b) and MEx (d) diagrams; 
and {\bf (iii)} a large fraction of SF galaxies on the blue diagram are in fact composites or AGNs 
according to the BPT diagram (b).

\begin{figure}[bh]
\epsscale{.5} \plotone{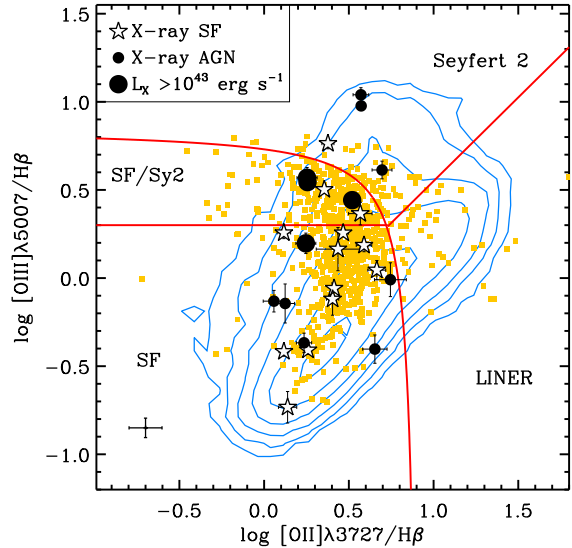}
\caption{
   The blue diagram, where AGNs are found above the solid curve
   and further separated between Seyfert 2 and LINER classes, as labeled.  
   Star-forming (SF) galaxies
   lie below the curve including a region where they are indistinguishable from Serfert 2 galaxies 
   (labeled as SF/Sy2).  In principle, this diagram is applicable out to $z \sim 1$.  However, the
   DEEP2 spectra used in this work span a narrow wavelength range and very few cover all three lines
   required in this diagram.  While performing well at selecting the X-ray starbursts (star symbols),
   this diagram misses the majority of X-ray AGNs (filled black circles) including the brighter ones
   (larger circles).
   Contours show the SDSS low-$z$ sample (evenly spaced on a logarithmic scale).
   Our intermediate redshift sample is superimposed (orange points) and, when available, the X-ray
   classification is marked with larger symbols [star symbols for X-ray starbursts; small (large) filled 
   circles for X-ray AGNs with $L_X < 10^{43}$~erg~s$^{-1}$ ($L_X > 10^{43}$~erg~s$^{-1}$)].
   (A color version of this figure is available in the online journal.)
   }\label{fig:blueX}
\end{figure}

The MEx diagram (Figure~\ref{fig:diagX}) is applicable to 2,812 galaxies
out to $z \sim 1$, whereas the emission lines in the blue diagram (Figure~\ref{fig:blueX}) 
are only available for 826 galaxies (29\%).  While 423 out of 531 (80\%) GOODS-N 
galaxies with \hb\ and \oiiilam\ also have a valid measurement of the \oiilam\ line, 
the situation is very different for EGS galaxies.  The DEEP2 spectra used for these 
galaxies span a more restricted range in wavelength and so only 403 among 2,536 objects 
(16\%) with valid \oiiilam\ and \hb\ also have a valid \oiilam\ line flux.

The star-forming region of the blue diagram (Figure~\ref{fig:blueX}) accounts for 
10/13 (77\%) of the X-ray starbursts but also for 6/11 (55\%) of the X-ray AGNs. The Seyfert 2 
region only accounts for 3/11 (27\%) of the X-ray AGNs.  The smaller number of galaxies on
the blue diagram hinder a quantitative comparison with the MEx diagram.  Nevertheless, it 
appears that the former is less reliable at selecting X-ray AGNs.


\bibliographystyle{apj}
\bibliography{paper1_bib}

\begin{deluxetable}{lcccccc}
\tablecolumns{7}
\tablecaption{Demographics of the MEx Diagram}
\tablehead{
   \colhead{BPT type}  &
   \colhead{MEx-AGN} &
   \colhead{(\%)}  &
   \colhead{MEx-Interm.\tablenotemark{a}} & 
   \colhead{(\%)}   &
   \colhead{MEx-SF} & 
   \colhead{(\%)}}
\startdata
BPT-SF        &  1465     &  6.0   &  9153   &  51.7 &  64243  &  94.4  \\
BPT-composite &  9782     &  40.0  &  8468   &  47.8 &  3760   &   5.5  \\
BPT-AGN       &  13212    &  54.0  &    90   &   0.5 &    31   &  0.04  \\
\hline
All           &  24459    &  100   &  17711  &  100  &  68034  &  100   
\tablenotetext{a}{~ MEx-intermediate region: between the two curves defined 
                    by Equations~\ref{eq:mexline} and \ref{eq:mexlower}.}
\enddata
\label{tab:diag}
\end{deluxetable}

\begin{deluxetable}{lccccccccl}
\tabletypesize{\scriptsize}
\tablecolumns{10}
\tablewidth{500pt}
\tablecaption{X-ray Stacking of Non-Detections}
\tablehead{
   \colhead{Sample}  &  \colhead{N\tablenotemark{a}} & \colhead{$<z>$} & \colhead{Soft\tablenotemark{b}}  & \colhead{Soft}  & \colhead{Hard\tablenotemark{b}}  & \colhead{Hard} & \colhead{HR\tablenotemark{c}}  & \colhead{$\Gamma$\tablenotemark{d}} & \colhead{Comment} \\
& & &  & \colhead{$\sigma$} &  & \colhead{$\sigma$} & & & 
}
\startdata
\multicolumn{10}{l}{\bf (i) MEx Diagram Selection (see \S\ref{sec:Xstack})} \\
P(AGN)$\geq$50\%        &  22 & 0.70 $\pm$ 0.18 & 1.61 $\pm$  0.54 & 4.6 & 1.61 $\pm$ 0.81 & 2.9  &  0.00 &   0.8  & Many absorbed AGNs \\
P(SF)$>$50\%         &  25 & 0.60 $\pm$ 0.19 & 2.20 $\pm$  0.54 & 6.3 & 1.49 $\pm$ 0.81 & 2.7  & -0.19 &   1.2  & Star-forming galaxies \\
                     &     &                 &                  &     &                 &      &       &        & and some absorbed AGNs \\
P(AGN)$\geq$30\%        &  34 & 0.72 $\pm$ 0.17 & 1.31 $\pm$  0.45 & 4.4 & 1.67 $\pm$ 0.69 & 3.6  &  0.12 &   0.8  & Many absorbed AGNs \\
P(SF)$>$70\%         &  13 & 0.54 $\pm$ 0.15 & 2.78 $\pm$  0.77 & 5.8 & 1.03 $\pm$ 1.10 & 1.4  & -0.46 &   1.7  & Mostly star-forming galaxies \\
                     &     &                 &                  &     &                 &      &       &        &  \\
30\%$\leq$P(AGN)$<$50\% &  12 & 0.77 $\pm$ 0.14 & 1.56 $\pm$  0.77 & 3.1 & 2.1  $\pm$ 1.2  & 2.6  &  0.15 &   0.6  & Many absorbed AGNs \\
                     &     &                 &                  &     &                 &      &       &        &  \\
\multicolumn{10}{l}{\bf (ii) Absorbed AGN Selection (based on $T \equiv L_{2-10keV}/L_{\oiiilam}$, see \S\ref{sec:Xstack_Cthik})} \\
log($T$)$<$0.25\tablenotemark{e}      &  13 & 0.71 $\pm$ 0.19 & 1.72 $\pm$  0.68 & 3.9 & 2.7  $\pm$ 1.0  & 4.0  &  0.22 &   0.4  & Very absorbed AGNs \\
\enddata
\tablenotetext{a}{~ Number of galaxies in the X-ray stack.}
\tablenotetext{b}{~ Count rate in units of 10$^{-6}$ count s$^{-1}$.}
\tablenotetext{c}{~ Hardness Ratio $\equiv$ (H-S)/(H+S), where H and S are the 
X-ray counts in the hard (2-8~keV) and soft (0.5-2~keV) bands.}
\tablenotetext{d}{~ Effective Photon Index.}
\tablenotetext{e}{~ Also required P(AGN)$>$30\% and $\log(M_{\star}[M_{\sun}])>10.2$ as selection criteria.}
\label{tab:X}
\end{deluxetable}

\end{document}